\documentclass{acm_proc_article-sp}

\usepackage[T1]{fontenc}

\usepackage{amsmath}
\usepackage{comment}
\usepackage{amssymb}
\usepackage{graphicx}
\usepackage{url}
\usepackage{times}
\usepackage{balance}
\usepackage[textsize=scriptsize]{todonotes}
\usepackage{xspace}
\usepackage{enumitem}
\usepackage{array}
\usepackage{capt-of}
\usepackage{subfigure}
\usepackage{pdflscape}
\usepackage{listings}
\definecolor{black}{rgb}{1,1,1}
\definecolor{olivegreen}{rgb}{0.2,0.8,0.5}
\definecolor{blue}{rgb}{0.1,0.1,0.8}
\definecolor{grey}{rgb}{0.5,0.5,0.5}
\lstdefinelanguage{ttl}{
sensitive=true,
morecomment=[l][\color{grey}]{@prefix},
morecomment=[l][\color{grey}]{@de},
morecomment=[l][\color{grey}]{@en},
morecomment=[l][\color{grey}]{\^\^},
morecomment=[l][\color{olivegreen}]{\#},
morecomment=[l][\color{blue}]{<http},
morestring=[b][\color{blue}]\",
}

\newcommand{\dyldo}{DyLDO\xspace}
\newcommand{\code}[1]{\textsf{\small #1}\xspace}
\newcommand{\codeSmall}[1]{\textsf{\scriptsize #1}\xspace}

\makeatletter
\def\@copyrightspace{\relax}
\makeatother

\begin{document}

\title{TermPicker: Enabling the Reuse of Vocabulary Terms by Exploiting Data from the Linked Open Data Cloud}
\subtitle{An Extended Technical Report}

\numberofauthors{3}

\author{
\alignauthor
Johann Schaible\\
       \affaddr{GESIS - Leibniz Institute for the Social Sciences}\\
       \affaddr{Germany}\\
       \email{johann.schaible@gesis.org}
\alignauthor
Thomas Gottron\\
       \affaddr{WeST - Institute for Web Science and Technologies}\\
       \affaddr{Germany}\\
       \email{gottron@gmail.com}
\alignauthor Ansgar Scherp\\
       \affaddr{ZBW -- Leibniz Information Centre for Economics, Germany}\\
       \affaddr{Knowledge Discovery, Kiel University, Germany}\\
       \email{a.scherp@zbw.eu}
       \email{asc@informatik.uni-kiel.de}
}

\date{}

\maketitle

\begin{abstract}
Deciding which vocabulary terms to use when modeling data as Linked Open Data (LOD) is far from trivial.
Choosing too general vocabulary terms, or terms from vocabularies that are not used by other LOD datasets, is likely to lead to a data representation, which will be harder to understand by humans and to be consumed by Linked data applications.
In this technical report, we propose \emph{TermPicker}: a novel approach for vocabulary reuse by recommending RDF types and properties based on exploiting the information on how other data providers on the LOD cloud use RDF types and properties to describe their data.
To this end, we introduce the notion of so-called \emph{schema-level patterns} (SLPs).
They capture how sets of RDF types are connected via sets of properties within some data collection, e.g., within a dataset on the LOD cloud.
TermPicker uses such SLPs and generates a ranked list of vocabulary terms for reuse.
The lists of recommended terms are ordered by a ranking model which is computed using the machine learning approach Learning To Rank (L2R).
TermPicker is evaluated based on the recommendation quality that is measured using the Mean Average Precision (MAP) and the Mean Reciprocal Rank at the first five positions (MRR@5).
Our results illustrate an improvement of the recommendation quality by $29-36\%$ when using SLPs compared to the beforehand investigated baselines of recommending solely popular vocabulary terms or terms from the same vocabulary.
The overall best results are achieved using SLPs in conjunction with the Learning To Rank algorithm \emph{Random Forests}.
\end{abstract}

\section{Introduction}
When modeling Linked Open Data (LOD), engineers employ Resource Description Framework (RDF) vocabularies to represent their data as LOD.
An RDF vocabulary is a collection of (unique) vocabulary terms, i.e., RDF types (also referred to as ''classes``) and properties, that represent a model about a certain domain.
It is considered best practice to choose RDF types and properties from existing vocabularies, i.e., reuse vocabulary terms, before defining proprietary terms to create a LOD model.
This reduces heterogeneity in the data representation by generating some ontological agreement with other data providers~\cite{series/synthesis/2011Heath}.
However, finding vocabulary terms that are \emph{appropriate} for reuse is far from trivial.
Prominent services, such as the, Linked Open Vocabularies catalog (LOV)~\cite{LOV}, vocab.cc~\cite{vocab-cc}, and others, can be used to find specific RDF types and properties based on string search.
LOV also provides an opportunity to use SPARQL for exploiting T-Box information on vocabularies and their terms, i.e., 
Linked Data engineers can search for equivalent RDF types or properties via \code{owl:equivalentClass} or \code{owl:equivalentProperty}, for sub-classes and sub-properties via \code{rdfs:}\code{subClass} or \code{rdfs:subProperty}, or for other relations between vocabulary terms which are defined within the vocabularies. 
However, these services do not exploit any A-Box information, unless a vocabulary is pointing to datasets that use the vocabulary.
Services like LODstats~\cite{auer2012lodstats} go a step further and use A-Box information to provide detailed statistics on the usage of vocabularies and vocabulary terms.
However, none of these services provide information on how data providers on the LOD cloud combine the RDF types and properties from the different vocabularies to model their entire dataset.

Thus, selecting appropriate vocabulary terms for reuse can still be time-consuming, if one intends to reuse terms that other LOD providers use for publishing similar data. 
One has to identify such LOD providers among the amount of different datasets on the LOD cloud and examine their data on instance-level, i.e., browse through resources and examine of which RDF type they are and which outgoing properties they have.

In this paper, we introduce \emph{TermPicker}:\footnote{This is a preliminary URL for the review process: \url{http://bit.ly/termpicker-eval}, last access 9/14/15}
a novel vocabulary term recommendation approach enabling the reuse of vocabulary terms by exploiting already published datasets on the LOD cloud.
It provides Linked Data engineers a possibility to choose RDF types and properties used by other LOD providers in a manner that is similar to the engineers' needs.
\begin{figure*}[t!]
\centering
\includegraphics[width=0.9\textwidth]{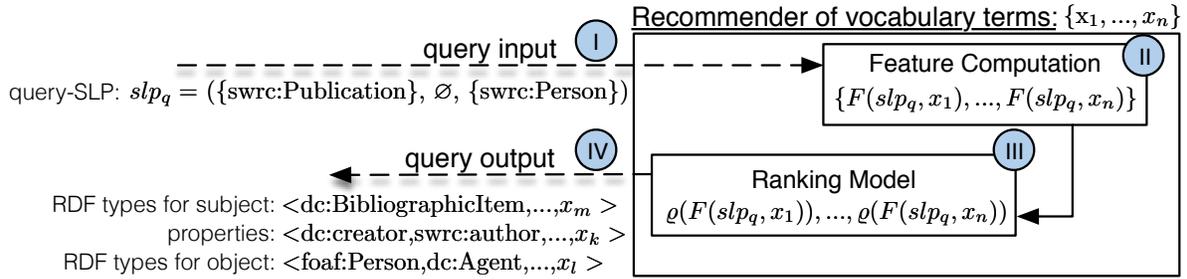}
\caption{\textit{Example.}
A Linked Data engineer models data as LOD illustrating publications and a persons, who are the corresponding creator of a publication.
She decided to reuse the SWRC vocabulary and has already chosen to use \codeSmall{swrc:Publication} and \codeSmall{swrc:Person}.
TermPicker uses this information and provides her with RDF vocabulary term recommendation from other vocabularies, such as FOAF, which were used by other LOD providers along with the chosen vocabulary terms.
In detail, Term\-Picker's input is the query-SLP $slp_q = (\{\code{swrc:Publication}\}, \varnothing, \{\code{swrc:Person}\})$ (step~(I)).
In step~(II), the query-SLP is extended by a recommendation candidate $x_i$ from the set $\{x_1,...,x_n\}$ of all terms published on the LOD cloud, and five features are calculated for each extended query-SLP.
The resulting feature values $F(slp_q, x_i)$ are used by the ranking model in step~(III) to order all vocabulary term recommendations from most to least appropriate, before providing the ranked lists as output in Step~(IV).
}
\label{fig:loverWorkflow}
\end{figure*}

To leverage the information how other LOD providers modeled their data, one needs to induce some structural information about which vocabulary terms were used to model entities and their relationships.
In this paper, the structural information is captured by so-called \emph{schema-level patterns} (SLPs).
A schema-level pattern is a tuple describing the connection between two sets of RDF types via a set of properties.
For example, the SLP
\begin{equation*}
\begin{aligned}
(\{\code{swrc:Publication}\}, \{\code{dc:creator}\}, \{\code{foaf:Person}\})
\end{aligned}
\end{equation*}
specifies that within one LOD collection (e.g. a dataset on the LOD cloud) resources of RDF type \code{swrc:Publication} are connected to other resources of RDF type \code{foaf:Person} via the property \code{dc:creator}.
The input for TermPicker is such an SLP that is specified by the user, i.e. the query-SLP $slp_q$.
TermPicker aims at extending the query-SLP by recommending additional vocabulary terms, which are used in other SLPs, which are calculated from existing datasets on the LOD cloud, and that are similar to $slp_q$.

The ranking of the recommendation candidates, i.e., vocabulary terms extracted from vocabularies published on the LOD cloud, is computed based on five features.
Three of the five features represent the \emph{popularity} of the recommendation candidate, i.e., how many data providers on the LOD cloud use the candidate, how many providers use the candidate's vocabulary, and what is the total number of occurrences of the candidate on the LOD cloud.
The fourth feature specifies if the recommendation candidate is from a vocabulary that is already used in the query-SLP $slp_q$.
Finally, the fifth feature is the so-called ``SLP-feature''.
It calculates the number of SLPs computed from datasets on the LOD cloud, which contain all terms from $slp_q$ as well as the recommendation candidate.
In other words, the SLP-Feature investigates whether other data providers on the LOD cloud have used the recommended term in a similar SLP to $slp_q$, i.e., in a similar manner.
The output is a set of three lists of vocabulary terms containing RDF types for resources and properties connecting these resources.
These lists are ordered by a ranking model, which is induced from some training data using the machine learning approach \emph{Learning To Rank} (L2R).
Learning To Rank is a family of supervised learning algorithms to establish a ranking over a set of items, in our case vocabulary terms, by observing a general coherence between the utilized features and the relevance of an item.

Figure~\ref{fig:loverWorkflow} illustrates TermPicker's general workflow and its components, such as the computation of the features and the ranking model $\varrho$.
Let us assume, a Linked Data engineer wants to model some data as LOD illustrating publications and each publication's creator.
She decided to reuse the SWRC\footnote{\url{http://swrc.ontoware.org/ontology}, last access 12/01/15} vocabulary and has already chosen to use \code{swrc:Publication} and \code{swrc:Person}.
In a first step, TermPicker receives the input in form of the query-SLP $slp_q = (\{\code{swrc:Publication}\}, \varnothing, \{\code{swrc:Person}\})$ ($\varnothing$ denotes an empty set).
TermPicker uses this information and provides the engineer with RDF vocabulary term recommendation from other vocabularies, such as FOAF\footnote{\url{http://xmlns.com/foaf/0.1/}, last access 12/01/15}, which were used by other LOD providers along with the chosen vocabulary terms.
To this end, $slp_q$ is extended with a recommendation candidate from a set of all vocabulary terms $\{x_1,...,x_n\}$ that are published on the LOD cloud, and the five features introduced beforehand are computed for the extended query-SLP.
The ranking model in the third step establishes three ranked lists of vocabulary terms that represent TermPicker's output.
One list contains RDF type recommendations for the resources in subject position, another one contains the RDF type recommendations for resources in object position, and the third one comprises recommendations of properties to connect these resources.
As these recommendations contain RDF types and properties from other vocabularies, the engineer is helped in finding also equivalent terms, which might better suit the engineer's need, e.g., using \code{foaf:Person} instead of \code{swrc:Person}.

We conduct a 10-fold leave-one-out evaluation to measure Term\-Picker's recommendation quality in different situations, in which one needs to select a vocabulary term for reuse.
The recommendation quality is assessed using the Mean Average Precision (MAP) and the Mean Reciprocal Rank at the first five positions (MRR@5).
As \emph{gold standard}, we do not rely on human judgment, but rather use an automated held-out approach, i.e., before providing TermPicker with a query-SLP, we randomly extract several terms from this SLP, and solely the extracted terms are considered relevant; each other recommended term is considered irrelevant.
We perform such an evaluation using data from the Billion Triple Challenge 2014~\cite{btc2014} as well as from the \dyldo seed-list~\cite{dyldo} dataset.
The query-SLPs for training and testing the ranking model are computed from ten different pay-level domains (PLDs), which have a relatively high ratio between reused vocabulary terms and all terms describing the data.
The triples and the calculated SLPs from the remaining PLDs represent the datasets that are already published on the LOD cloud.
The calculated SLPs from nine PLDs are used to train the ranking model and the calculated SLPs from one PLD are used to validate the ranking model.
As the SLPs are computed from real-world data, they vary by different vocabulary terms and by the number of contained vocabulary terms.
To evaluate different ranking models, we use the L2R algorithms contained in the RankLib\footnote{\url{http://sourceforge.net/p/lemur/wiki/RankLib/}, last access 9/14/15} library, which provides an entire framework to train and evaluate diverse ranking models.
Summarizing, the main contributions of this paper are:
\begin{enumerate}[label=(\roman*),leftmargin=0.8cm,itemsep=1ex]
	\item  Evaluation of the diverse Learning To Rank algorithms contained in the RankLib library that are used to calculate a ranking model for TermPicker's recommendations.
	\item Evaluation of the SLP-feature's impact on the recommendation quality by comparing its recommendations to the baselines of recommending solely popular vocabulary terms and recommending terms from an already used vocabulary~\cite{2014SchaibleESWC, lodipublishing}.
	\item Evaluation of the different recommendations regarding whether to choose an RDF type for resources in subject position of a triple, an RDF type describing resources in object position, or to pick a property, as this reflects different real-world LOD modeling scenarios~\cite{Noy101}.
\end{enumerate}

The paper is structured as follows:
Section~\ref{sec:SLP} describes the notion of schema-level patterns in detail and depicts how they are computed from RDF triples.
Section~\ref{sec:approach} illustrates the general workflow of the proposed recommendation approach including a detailed description of the five features and a brief introduction to L2R.
The evaluation of the proposed approach is described in Section~\ref{sec:eval}, whereas the results of the evaluation are illustrated in Section~\ref{results}.
TermPicker and the evaluation results are discussed in Section~\ref{discussion}.
The related work is discussed in Section~\ref{relWork}, in which we also differentiate TermPicker's approach to existing tools and services, before we conclude the paper.

\section{Schema-Level Patterns (SLPs)}\label{sec:SLP}
When reusing vocabularies with the goal to preferably result in some ontological agreement in data representation, one must investigate how other Linked Data providers modeled their data.
Investigating solely the specification or documentation of vocabularies does not provide such information.
To know which properties are used to connect resources of specific RDF types, existing datasets published on the LOD cloud must be investigated on instance level, i.e., one must browse through the data.
This can be very time consuming, specifically as the number of datasets on the LOD cloud is rising.

To alleviate the situation, we introduce the notion of schema-level patterns (SLPs).
They illustrate how the resources from a dataset on the LOD cloud are connected.
For example, the schema-level pattern
\begin{equation}
	\begin{aligned}
		slp = (&\{\code{foaf:Person}, \code{dbo:ChessPlayer}\}, \\
	 			&\{\code{foaf:knows}\}, \{\code{foaf:Person}, \code{dbo:Coach}\}) \label{eq:computedSlps}		
	\end{aligned}
\end{equation}
illustrates that resources of types \code{foaf:Person} and \code{dbo:ChessPlayer} are connected to resources of types \code{foaf:Person} and \code{dbo:Coach} via the property \code{foaf:knows}.
Such SLPs can be calculated from existing data sets on the LOD cloud, i.e., the SLPs are calculated based on an RDF triple representation, such as N3\footnote{\url{http://www.w3.org/TeamSubmission/n3/}, last access 12/01/15}, Turtle\footnote{\url{http://www.w3.org/TR/turtle/}, last access 12/01/15}, or others.
The SLP in equation (\ref{eq:computedSlps}) is calculated from the fictive RDF triples in Listing~\ref{lst:slpInstanceTriple}.
\begin{lstlisting}[float=t, breaklines=true,language=ttl,basicstyle=\scriptsize\ttfamily, xleftmargin=.15in, label=lst:slpInstanceTriple, 
caption=\footnotesize{\textbf{Fictive RDF triples in Turtle syntax.} The RDF triples specify that a resource of types \codeSmall{Person} and \codeSmall{ChessPlayer} knows a resource of types \codeSmall{Person} and \codeSmall{Coach}}]
@prefix rdf: <http://www.w3.org/1999/02/22-rdf-syntax-ns#>
@prefix foaf: <http://xmlns.com/foaf/0.1/>
@prefix dbo: <http://dbpedia.org/ontology/>

<http://ex1.org/sports_001> 
	 rdf:type foaf:Person;
	 rdf:type dbo:ChessPlayer;
	 foaf:knows <http://ex1.org/employee_002>.
	 
<http://ex1.org/sports_002> 
	 rdf:type foaf:Person;
	 rdf:type dbo:Coach.
\end{lstlisting}

SLPs provide an easy to use possibility for investigating how other data providers on the LOD cloud have modeled their data without having to look into the data itself.
Thus, choosing vocabulary terms that are recommended based on SLPs will eventually result in an ontological agreement in data representation.

In the following we define schema-level patterns formally (cf. Section~\ref{sec:slpIntroduction}) and describe how they can be computed from existing LOD sources in Section~\ref{sec:computedSLPs}.  

\subsection{Formal Definition of Schema-Level Patterns}\label{sec:slpIntroduction}
For a better overview, the most important variables used to define SLPs are enlisted in Table~\ref{tab:secTwoVars}.
\begin{table}[t!]
	\caption{Tabular overview of the variables that are used and explained in Section~\ref{sec:slpIntroduction} and Section~\ref{sec:computedSLPs}} \label{tab:secTwoVars}
\begin{tabular}{p{1cm} p{6.6cm}}
		Variable 	& Definition \\
		\hline \hline
		$\mathbb{V}$ & Set of all vocabularies on the LOD cloud \\
		$\mathbb{T}$ & Set of all RDF types from all vocabularies in $\mathbb{V}$ \\
		$\mathbb{P}$ & Set of all properties all vocabularies in $\mathbb{V}$ \\
		$slp$ & A schema-level pattern with $slp = (sts, ps, ots)$ \\
		$sts$ & Subject type set with $sts \in \mathcal{P}(\mathbb{T})$: RDF types describing a resource in subject position of a triple \\
		$ots$ & Object type set with $ ots \in \mathcal{P}(\mathbb{T})$: RDF types describing a resource in object position of a triple \\
		$ps$ & Property set with $ ps \in \mathcal{P}(\mathbb{P})$: properties interlinking resources of types in $sts$ and $ots$ \\
		\hline
		$\mathbb{DS}$ & The set of datasets that are published on the LOD cloud\\
		$G$ & A graph representing a dataset such that $G \in \mathbb{DS}$ \\
		$(s,p,o,c)$ & An RDF quadruple consisting of a subject, property, object, and a context URI where $G$ can be found  \\
		\hline		
	\end{tabular}
\end{table}

Let $\mathbb{V} = \{V_1, V_2, ... , V_n\}$ be the set of all vocabularies used by datasets on the LOD cloud.
Each vocabulary $V \in \mathbb{V}$ consists of vocabulary terms that are either an instance of \code{rdfs:Class} or \code{rdfs:Property}, such that $V = P_V \, \cup \, T_V$, where $P_V$ is the set of all properties $p$ and $T_V$ is the set of all RDF types $t$ in vocabulary $V$.
Accordingly, $\mathbb{T} = \bigcup_{V \in \mathbb{V}} T_V$ is the set of all RDF types and $\mathbb{P} = \bigcup_{V \in \mathbb{V}} P_V$ the set of all properties on the LOD cloud. 
The formal definition of an SLP is
\begin{equation}
\begin{aligned}
	slp \in \mathcal{P}(\mathbb{T}) \times \mathcal{P}(\mathbb{P}) \times \mathcal{P}(\mathbb{T})
\end{aligned}
\end{equation}
where $\mathcal{P}(\mathbb{T})$ is the power set of all RDF types and $\mathcal{P}(\mathbb{P})$ the power set of all properties on the LOD cloud.
Based on this, an SLP is a tuple
\begin{equation}
\begin{aligned}
	slp &= (sts, ps, ots)
\end{aligned}
\end{equation}
where $sts \in \mathcal{P}(\mathbb{T})$ is the set of RDF types describing resources in subject position of a triple, $ots \in \mathcal{P}(\mathbb{T})$ the set of RDF types describing resources in object position of a triple, and $ps \in \mathcal{P}(\mathbb{P})$ the set of properties interlinking the resources of types in $sts$ and $ots$.

To operate with SLPs, we define the two operators $\oplus$ and $\ominus$.
The commutative $\oplus$ operator combines two SLPs:
\begin{equation}
		slp_i  \oplus slp_j := (sts_i \cup sts_j, ps_i \cup ps_j, ots_i \cup ots_j)
\end{equation}
It can also be used for extending an SLP with a further vocabulary term by adding it either to the sets $sts$, $ps$, or $ots$.
In detail, the operator $\oplus_{sts}$ adds an RDF type to the set $sts$, operator $\oplus_{ots}$ adds a RDF type to $ots$ and the operator $\oplus_{ps}$ adds a property to the set of properties $ps$.
This is specifically useful for examining whether a query-SLP is used in combination with a recommendation candidate by other data providers on the LOD cloud.
The operation to remove terms from an SLP via the $\ominus$ is defined accordingly.
The operator $\ominus_{sts}$ removes an RDF type from the set $sts$, operator $\ominus_{ots}$ a RDF type from $ots$ and the operator $\ominus_{ps}$ removes a property from the set of properties $ps$.
An example for first removing a property from an SLP and subsequently extending the SLP with an RDF type for resources in object position would be as follows:
\begin{equation*}
	\begin{aligned}
		slp = &(\{\code{foaf:Person}\}, \{\code{dc:date}\}, \varnothing) \ominus_{ps} \code{dc:date} \\
			= &(\{\code{foaf:Person}\}, \varnothing, \varnothing) \\
		slp = &(\{\code{foaf:Person}\}, \varnothing, \varnothing) \oplus_{ots} \code{foaf:Image} \\
			=	&(\{\code{foaf:Person}\}, \varnothing, \{\code{foaf:Image}\})
	\end{aligned}
\end{equation*}

The relationship ``$\leq$'' between two schema-level patterns $slp_i$ and $slp_j$ illustrates that one SLP can be a \emph{subset} of another SLP.
It is defined as
\begin{equation}
\begin{aligned}
	slp_i \leq slp_j, \quad \text{iff} \quad &(sts_i \subseteq sts_j) \wedge (ps_i \subseteq ps_j) \\ 
															&\wedge (ots_i \subseteq ots_j)
\end{aligned}
\end{equation}
and illustrates that $slp_j$ contains more or at least as many vocabulary terms as $slp_i$.
The strict relation $slp_i < slp_j$ defines that at least one set $sts_i$, $ps_i$, or $ots_i$ is a proper subset of $sts_j$, $ps_j$, or $ots_j$, respectively.
Such a relation is useful for comparing two SLPs, especially to inspect whether a query-SLP in conjunction with a recommendation candidate is a subset of other SLPs calculated from datasets on the LOD cloud.

\subsection{Computing SLPs from Linked Open Data}\label{sec:computedSLPs}
Let $\mathbb{DS} = \{G_1, G_2, ... , G_m\}$ be the set of all data sources on the LOD cloud.
Hereby, $G$ denotes the graph of the data source and can be considered as a set of quadruples with
\begin{equation}
\begin{aligned}
		G = \{ (s, p, o, c) \, | \,  	&s \in \textit{URI} \cup \textit{BN}, p,c \in \textit{URI},  \\
										&o \in \textit{URI} \cup \textit{BN} \cup \textit{LIT}\}
\end{aligned}
\end{equation}
where \textit{URI} is a set of URI's, \textit{BN} a set of blank nodes, and \textit{LIT} a set of literals.
A triple consists of $s$, $p$, and $o$ with $s$ being the subject, $p$ being the property, and $o$ being the object of a triple.
The context URI $c$ specifies where graph $G$ can be found.
Function $\lambda (G) = \{slp_1, slp_2, ..., slp_k\}$ defines the set of SLPs that are computed from the according graph $G$.
The specification of $\lambda: \mathbb{DS} \rightarrow \mathbb{SLP}$ is
\begin{equation} 
\begin{aligned}  
\lambda (G) = \{ &(sts, ps, ots) \mid \exists \,  s,o : \\
							&\left(\forall \, t_s \in \textit{sts} : (s, \code{rdf:type}, t_s) \in G \right ) \wedge  \\
							&\left(\forall \, p \in \textit{ps} :  (s, p, o) \in G \right ) \wedge \\
							&\left(\forall \, t_o \in \textit{ots} : (o, \code{rdf:type}, t_o) \in G\right )\}
\end{aligned}
\end{equation}
Hereby, $\mathbb{SLP}$ is the joint set of schema-level patterns that are computed from each graph $G \in \mathbb{DS}$
\begin{equation}
	\begin{aligned}
	\mathbb{SLP} = \bigcup_{G \in \mathbb{DS}}\left( \lambda (G) \right)
	\end{aligned}
\end{equation}
An example for calculating an SLP from a graph $G$ is provided in Equation~(\ref{eq:computedSlps}), which illustrates a computed SLP from the data listed in Listing~\ref{lst:slpInstanceTriple}.

\section{Picking Vocabulary Terms using SLPs}\label{sec:approach}
Besides illustrating how resources of specific RDF types are connected to each other, schema-level patters can be used to recommend vocabulary terms for reuse.
TermPicker's input is an SLP, i.e., the query-SLP $slp_q$.
It is extended with a vocabulary term $x$ from the set of terms from all data sources on the LOD cloud ($x \in (\mathbb{T} \cup \mathbb{P})$).
These vocabulary terms are considered to be \emph{recommendation candidates}.
Subsequently, TermPicker compares the extended query-SLP to all SLPs in $\mathbb{SLP}$.
Each SLP $slp_i$ with
\begin{equation}
\begin{aligned}
slp_i \in \mathbb{SLP} \quad \text{and} \quad &(slp_q \oplus_{sts} x) \leq slp_i \, \vee \\
							&(slp_q \oplus_{ps} x) \leq slp_i \, \vee \\
							&(slp_q \oplus_{ots} x) \leq slp_i \, ,
\end{aligned}
\end{equation}
is an SLP that uses vocabulary term $x$ in combination with the terms in $slp_q$.
Thus, vocabulary term $x$ can be generally considered a \emph{good} recommendation candidate for reuse.
For providing meaningful recommendation candidates, the query-SLP must not be empty, i.e.,  $slp_q \neq (\varnothing, \varnothing, \varnothing)$, otherwise each term $x$ would be considered a good recommendation.
Also, for better readability of the paper, we generalize the extension of a query-SLP \linebreak $slp_q = (sts_q, ps_q, ots_q)$ by a term $x$ with 
\begin{equation}
\begin{aligned}
	slp_q \oplus x := sts_q \cup x \vee ps_q \cup x \vee ots_q \cup x
\end{aligned}
\end{equation}
that specifies: $slp_q$ is extended with a vocabulary term $x$ by adding term $x$ either to the set $sts_q$, $ps_q$, or $ots_q$.

However, considering solely the existence of SLPs from $\mathbb{SLP}$ that use a recommendation candidate $x$ in combination with the terms in $slp_q$, might not be sufficient to provide most reasonable recommendations.
To rank each recommendation candidate from \emph{most appropriate} to \emph{least appropriate}, one should also encounter the popularity of the recommendation candidate and whether it is from a vocabulary that is already used in the query-SLP~\cite{2014SchaibleESWC, lodipublishing}.
Thus, one must first define a set of features representing each of these aspects of the recommendation candidates.
A ranking model then puts the recommendation candidates in order by using these features.
Establishing a general ranking model based on observing coherences between the features manually is a challenging task.
Therefore, TermPicker utilizes a Learning To Rank (L2R) algorithm that observes such coherences in an automatic way.

In the following, we describe and explain each feature that is used to categorize a recommendation candidate as well as the features' computation in Section~\ref{sec:ProblemStatement}.
The machine learning approach Learning To Rank and how it is used to generate a ranking model for recommending vocabulary terms is briefly illustrated in Section~\ref{sec:l2r}.

\subsection{Feature Computation}\label{sec:ProblemStatement}
The set of features that categorize each recommendation candidate $x$ is enlisted in Table~\ref{tab:features}.
\begin{table}[t!]
	\caption{\textbf{Overview of the utilized features.} The features are computed for every recommendation candidate $x \in (\mathbb{T} \cup \mathbb{P})$} \label{tab:features}
\begin{tabular}{m{0.8cm} m{6.6cm}}
		\multicolumn{1}{c}{Feature} & \multicolumn{1}{c}{Definition} \\
		\hline \hline
		$f_1$ & Number of datasets on the LOD cloud using the recommendation candidate $x$ \\
		$f_2$ & Number of datasets on the LOD cloud using the vocabulary $V_x$ of recommendation candidate $x$ \\
		$f_3$ & Total number of occurrences of recommendation candidate $x$ on the LOD cloud  \\
		$f_4$ & Whether the recommendation candidate $x$ is from a vocabulary that is already used in query-SLP $slp_q$  \\
		$f_5$ & Number of SLPs in $\mathbb{SLP}$ that contain recommendation candidate $x$ in conjunction with $slp_q$ \\
		\hline		
	\end{tabular}
\end{table}
This set of features was derived from~\cite{2014SchaibleESWC}, which illustrated that the most common strategies and influencing factors to choose a vocabulary terms for reuse is its \emph{popularity} and whether or not it is from a vocabulary that is already used.
Features $f_1$ to $f_3$ represent the popularity of a vocabulary terms whereas feature $f_4$ specifies whether the recommended term is from a vocabulary that is is already used in the query-SLP.
Additionally, we introduce feature $f_5$ that calculates how many SLPs $slp_i \in \mathbb{SLP}$ exist with $slp_q \oplus x \leq slp_i$.
Each of these features represent some factor that an engineer might consider important in her vocabulary term choice.
However, none of the features encode the relevance of a recommendation candidate directly.
In Sections~\ref{sec:feature_1} to~\ref{sec:feature_5}, these five features are described in detail including the formalizations for their computation.

\subsubsection{Popularity (Features $f_1$ to $f_3$)}\label{sec:feature_1}
Feature $f_1$ comprises the number of datasets $G \in \mathbb{DS}$ on the LOD cloud using a recommendation candidate $x$.
It is calculated by examining whether the term $x$ is contained in an RDF triple/quadruple of a graph $G$.
\begin{equation}
	\begin{aligned} 
		f_1(x) =| \{ &G \, \mid \, (\exists \, (s,p,o,c) \in G: p = x) \, \vee \\
							&(\exists \, (s, \code{rdf:type}, o, c) \in G: o = x) \}|
	\end{aligned}
\end{equation}
Feature $f_2$ depicts the number of datasets on the LOD cloud using the vocabulary $V_x$ of the recommendation candidate $x$.
It is calculated similar to feature $f_1$, but it examines whether the vocabulary of term $x$ is used in a triple of graph $G$.
\begin{equation}
	\begin{aligned} 
		f_2(V_x) =| \{ &G \, \mid \, (\exists \, (s,p,o,c) \in G: p \in V_x  \, \vee \\
		&(o \in V_x \wedge p = \code{rdf:type}))\}|
	\end{aligned}
\end{equation}
The total number of occurrences of the recommendation candidate $x$ on the LOD cloud is calculated by feature $f_3$.
In contrast to the features $f_1$ and $f_2$, feature $f_3$ is calculated by counting each triple/quadruple, in which the vocabulary term $x$ is contained.
\begin{equation}
	\begin{aligned} 
		f_3(x) = \sum_{G \in \mathbb{DS}} |\{&(s,p,o,c) \in G \, \mid \, (p = x) \, \vee \\
		&(o = x \wedge p = \code{rdf:type})\}|
	\end{aligned}
\end{equation}

Combined, these three features define the popularity of a vocabulary term on a very fine-grained level.
Whereas the total number of occurrences of a recommendation candidate $x$ depicts its overall usage, the number of data sources using $x$ and its vocabulary specifies whether its usage is spread across many datasets on the LOD cloud or concentrates on only a few ones.
We do not normalize any of the feature values, but rather use the absolute values, as this ensures that valuable information would not be lost, i.e., normalizing the feature values in our L2R based evaluation setup could lead to false recommendations.

The benefit of reusing popular vocabulary terms is supposed to enable an easier consumption of the data, as many Linked Data consumption tools provide tailored support for popular vocabularies~\cite{series/synthesis/2011Heath}.
This is also backed up by the recommendations of the W3C when modeling LOD.\footnote{\url{http://www.w3.org/TR/ld-bp/#VOCABULARIES}, last access 12/12/15}
In addition, it makes the data more understandable for humans.
TermPicker makes use of these features, as they are also acknowledged by Linked Data practitioners in a survey on their strategies and influencing factors to reuse a vocabulary term or not~\cite{2014SchaibleESWC}.

\subsubsection{Same Vocabulary (Feature $f_4$)}
Feature $f_4$ indicates whether the vocabulary of a recommendation candidate $x$ is already contained in the query-SLP, i.\,e., \linebreak $slp_q = (sts_q, ps_q, ots_q)$.
The calculation returns a binary value, where $1$ denotes that the vocabulary of term $x$ is already used in $slp_q$, and $0$ if it is not contained in $slp_q$.
\begin{equation}
	\begin{aligned} 
		f_4(slp_q, x) = \left\{
			\begin{array}{l l}
				1 & \quad \text{if } \exists \, V : x \in V \wedge \\
					& \quad (sts_q \cup ps_q \cup  ots_q) \cap V  \neq \emptyset \\
				0 & \quad \text{else}
			\end{array} \right.
	\end{aligned}
\end{equation}
Reusing terms from the same vocabulary is considered as an important strategy not only in the survey on vocabulary reuse strategies described in~\cite{2014SchaibleESWC}, but specifically in certain domains such as the statistics domain.
There, it is accustomed to reuse primarily vocabulary terms from SKOS\footnote{\url{http://www.w3.org/2004/02/skos/}, last access 09/06/15} or XKOS\footnote{\url{http://rdf-vocabulary.ddialliance.org/xkos.html}, access 09/06/15}~\cite{lodipublishing}.
In other words, one might want to search for vocabularies covering the domain of interest and subsequently adapt RDF types and properties from those vocabularies for particular needs.
The reason for that: it seems quite likely that one specific domain vocabulary, such as SKOS, contains many RDF types or properties that can be reused for describing data from that specific domain.
Furthermore, reusing terms from the same vocabulary reduces the overload of too many different vocabularies and makes the data easier to understand for humans that are familiar with the domain specific vocabulary~\cite{2014SchaibleESWC}.

\subsubsection{The SLP-Feature (Feature $f_5$)}\label{sec:feature_5}
The SLP-feature is calculated based on a query-SLP $slp_q$ that is extended with a recommendation candidate $x$.
The extended query-SLP $slp_q \oplus x$ is compared to existing SLPs $slp_i \in \mathbb{SLP}$, in order to find SLPs with $(slp_q \oplus x) \leq slp_i$.
The number of SLPs $slp_i \in \mathbb{SLP}$ with $(slp_q \oplus x) \leq slp_i$ represents how often other datasets on the LOD cloud use vocabulary term $x$ in conjunction with the terms in $slp_q$.
\begin{equation}
	\begin{aligned} 
		f_5((slp_q &\oplus x) \,, \, \mathbb{SLP}) =  |\{slp_{i} \;| \; slp_q \oplus x &\leq slp_i \}|
	\end{aligned}
\end{equation}
Using recommendations based on this feature is likely to result in reducing heterogeneity in the data representation by relying on ontological agreement.
The more SLPs in $\mathbb{SLP}$ use the recommendation candidate $x$ in such a \emph{similar} way, the more appropriate does it seem to reuse this term in order to eventually result in some ontological agreement.

\subsection{Learning to Rank}\label{sec:l2r}
Combined, features $f_1$ to $f_5$ describe each recommendation candidate $x$ in a unique way.
However, it remains unclear how these features can be used to provide a ranked list of recommendations.
The feature values for each recommendation candidate might vary a lot, as in the following fictive example:
\begin{itemize}
	\item $(slp_q \oplus x_1) = $ $f_1$: 7, $f_2$: 9, $f_3$: 20, $f_4$: 1, $f_5$: 4 
	\item $(slp_q \oplus x_2) = $ $f_1$: 3, $f_2$: 3, $f_3$: 5, $f_4$: 0, $f_5$: 6 
	\item $(slp_q \oplus x_3) = $ $f_1$: 10, $f_2$: 30, $f_3$: 80, $f_4$: 0, $f_5$: 2 
	\item $(slp_q \oplus x_4) = $ $f_1$: 4, $f_2$: 20, $f_3$: 29, $f_4$: 1, $f_5$: 0 
\end{itemize}
Immediately, the question arises which of these four recommendation candidates can be considered the most appropriate term for reuse.
To rank these terms from most to least appropriate, one must observe a general coherence between the features and the relevance of each recommendation candidate.
However, observing such a coherence manually can be quite difficult.
Rather, it must be observed in an automatic way to learn the feature's impact on the quality of the recommendations.

In order to address this challenge, TermPicker makes use of the machine learning approach ``Learning To Rank'' (L2R).
Learning to rank refers to a class of supervised machine learning techniques for inducing a ranking model~\cite{liu2009learning, hang2011short}.
In detail, a ranking model $\varrho$ allows for determining relevant and irrelevant items for a given information need.
In our case, an information need corresponds to the query-SLP $slp_q$.
The relevant and irrelevant items correspond to the recommendation candidates $x \in \mathbb{T} \cup \mathbb{P}$.
The ranking model $\varrho$ is derived from some training data by observing the mentioned general coherence between the feature values and the relevance of a recommendation candidate.
Ideally, the derived ranking model lists all relevant vocabulary terms high and before less relevant or irrelevant vocabulary terms.

Formally, the ranking model ($\varrho(F(slp_q, x))$) calculates a ranking score for the recommendation candidate $x$, where $F(slp_q, x)$ denotes the calculation of features $f_1$ to $f_5$ for the extended query-SLP $slp_q$ by the recommendation candidate $x$.
This way, each recommendation candidate $x \in \mathbb{T} \cup \mathbb{P}$ can be ranked based on the ranking score in descending order.
To establish such a ranking model, one needs training data to derive a general coherence between the feature values and the relevance of a recommended term.
In our case, the training data is a set of query-SLPs with existing relevance information on each recommendation candidate.
It contains SLPs such as 
$$slp_q = (\{\code{swrc:Publication}\}, \varnothing, \{\code{foaf:Agent}\})$$
with the relevance information that e.g. for recommending properties solely the terms \code{dc:creator} and \code{swrc:author} are considered as relevant.
Using this information, an L2R algorithm iterates through the training data to detect the beforehand mentioned coherence between the feature values and the relevance, such that the relevant terms get ranked as high as possible.
This way, the learned ranking model can be used in new and previously unknown situations with new and unknown query-SLPs. 
For example, a query-SLP that was not part of the training set using terms from the Creative Commons\footnote{\url{http://creativecommons.org/ns#}, last access 09/06/15} ontology and from an ontology for managing presentations at W3C\footnote{\url{http://www.w3.org/2004/08/Presentations.owl#}, last access 09/06/15}
$$slp_q = (\{\code{cc:Work}\}, \{\code{w3:presenter}\}, \varnothing) $$
can get recommendations, such as the RDF types \code{foaf:Person} and/or \code{dc:Agent} to reuse for resources in object position. 

L2R algorithms are categorized in three different groups according to their method for learning a ranking model~\cite{liu2009learning}:
(A)~\emph{point-wise} L2R algorithms, 
(B)~\emph{pair-wise} L2R algorithms, and 
(C)~\emph{list-wise} L2R algorithms.
A point-wise approach ranks vocabulary terms directly by allocating a ranking score to each recommendation candidate individually.
Pair-wise methods rank vocabulary terms solely in a given pair of two recommendation candidates.
This way, a term is considered a better recommendation compared to the terms in a lower ranking position.
List-wise approaches rank recommendation candidates by optimizing the quality measure of the result list, such as the Mean Average Precision (MAP).
They examine which coherence between the features provides the highest measure, e.g., the highest MAP value, and use the derived ranking model assuming the quality measure is as high in new situations.

In particular, the pair-wise and list-wise approaches have demonstrated good performance in generic ranking scenarios~\cite{busa2012apple}.
However, it is of interest for our use-case to determine which of the approaches, i.e., point-wise, pair-wise, or list-wise, perform better in our setting of recommending vocabulary terms for reuse.

\section{Evaluation}\label{sec:eval}
The proposed approach is evaluated using a 10-fold leave-one-out evaluation.
Each fold comprises a \emph{training set} to induce the ranking model, a \emph{test set} to evaluate the ranking model, and a set representing datasets that are already published on the LOD cloud to calculate features $f_1$ to $f_5$.
We investigate different ranking models and thus TermPicker's recommendation quality based on the aspects that depict the main contribution of this paper:
\begin{enumerate}[label=(\roman*),leftmargin=0.8cm,itemsep=1ex]
	\item Comparison of all Learning To Rank algorithms contained in the RankLib library that provides a framework for inducing and evaluating a ranking model. The three most competitive Learning To Rank algorithms are examined in detail, i.e., in our evaluation these three algorithms were \emph{Coordinate Ascent}~\cite{CoordAscent}, \emph{LambdaMART}~\cite{wu2010adapting}, and \emph{Random Forests}~\cite{RanForr}.
	\item Comparison of using the SLP-feature ($f_5$) to using features $f_1 - f_3$ (baseline of reusing only popular vocabulary \linebreak terms)~\cite{2014SchaibleESWC} and to using features $f_1 - f_4$ (baseline of reusing popular vocabularies from the same vocabulary)~\cite{lodipublishing} to investigate the impact of the SLP-feature on the recommendation quality.
	\item Comparison of recommending RDF types for resources in subject position of a triple, RDF types describing resources in object position, and recommending properties, as this reflects different real-world LOD modeling scenarios~\cite{Noy101}.
\end{enumerate}
The recommendation quality is measured using the Mean Average Precision (MAP) and the Mean Reciprocal Rank at the first five positions (MRR@5).

In the following, Section~\ref{sec:experimentDesign} describes the evaluation design in detail.
It is illustrated how the relevance of a recommendation candidate is defined, in order to enable the L2R algorithm to learn the ranking model.
In Section~\ref{sec:evalData} it is explained which data was used for the evaluation as well as how it was split to train and evaluate the ranking model.
It also includes statistics on the data and the ten folds.
Finally, we formalize the quality measures MAP and MRR@5 to illustrate how the recommendation quality was calculated.

\subsection{Evaluation Design}\label{sec:experimentDesign}
TermPicker's recommendations are evaluated by simulating a search for an appropriate vocabulary term that can be reused.
Thus, the training set and test set, which are used to induce and evaluate the ranking model, are disjunct sets of distinct SLPs.
These SLPs are used as input for TermPicker.
However, before providing TermPicker with these SLPs as input, one or more random vocabulary terms are extracted from that SLP using the $\ominus$ operator.
These extracted terms determine the set of \emph{relevant} recommendation candidates, as they are the ones that have been initially used.
All other recommendation candidates that are not contained in the set of the extracted terms are considered as irrelevant recommendations.
This way, for each query-SLP, the ranking model is provided 
(a)~a set of recommendation candidates, 
(b)~five feature values categorizing each recommendation candidate, and 
(c)~the relevance of each recommendation candidate.
The L2R algorithm uses this information and observes a general coherence between the feature values and the relevance of a recommendation~\cite{hang2011short}.

For example, given an SLP $slp_j$ from the training or test set with 
\begin{equation*}
	\begin{aligned}
		slp_j = (&\{\code{swrc:Publication}\}, \{\code{swrc:author}\}, \{\code{swrc:Person}\})		
	\end{aligned}
\end{equation*}
the property \code{swrc:author} is randomly extracted via the $\ominus_{ps}$ operator.
\begin{align*}
		slp_q 	&= slp_j \ominus_{ps} \code{swrc:author} \\
					&= (\{\code{swrc:Publication}\}, \varnothing, \{\code{swrc:Person}\}) 
\end{align*}
The query-SLP $slp_q$ is now provided as input for TermPicker.
The output is a set of vocabulary terms, including a set of properties.
The previously extracted property \code{swrc:author} is considered a relevant recommendation, as it was initially used in $slp_j$.
Every other recommendation is considered irrelevant, as these terms were not used in $slp_j$.
This makes it possible to induce and evaluate a ranking model by interpreting a ranked list of recommendations
$$<\code{dc:date}, \code{dc:title}, \code{\textbf{swrc:author}},...>$$
in the following way: the first two recommendations are irrelevant, and the first relevant recommendation is at the third rank of the result list.

Such an evaluation can be performed fully automatically reflecting many different real-life scenarios.
Human assessment whether a recommendation is relevant or not is not required.
This helps drastically to establish a first generalized ranking model using a lot of data.
Relying on human judgment would be very time consuming and difficult, as the manual assessment would take a lot of time and one would need many different domain experts, in order to correctly judge every recommendation candidate.
The real-life scenarios are represented by the many different query-SLPs.
Each query-SLP represents the Termpicker's input provided by the engineer, and the previously extracted term represents what the engineer is looking for.
Every recommendation candidate is assigned its feature values.
Sometimes the previously extracted term is used by other LOD providers in conjunction with the query-SLP and sometimes not, which is reflected by the SLP feature value.
Thus, the SLP feature is only an indicator that a recommendation candidate might be relevant, and therefore, the automatic evaluation provides every aspect in order to evaluate how much influence the SLP feature actually has on the recommendation quality.

\subsection{Datasets for the Evaluation}\label{sec:evalData}
\begin{table*}[t!]
	\caption{PLDs that were selected as test and training in the evaluations. The selection was based on $C1$ (PLDs that provided the highest number of distinct vocabulary terms) and $C2$ (PLDs with the highest ratio between the reused vocabulary terms and all RDF types and properties). The left half of the table shows the selected PLDs from the \dyldo dataset, whereas the right half shows the selected PLDs from the BTC 2014 dataset} \label{tab:evalDataStats}
	\centering
\begin{tabular}{r  r  r  r || r  r  r  r}
		\multicolumn{4}{c}{\dyldo} & 	\multicolumn{4}{c}{BTC 2014} \\
		\hline
		PLD 	& $(C1)$ & $(C2)$ & \# of SLPs & PLD & $(C1)$ & $(C2)$ & \# of SLPs \\
		\hline \hline
		kasei.us & 	100	&	1.00	&	121	&	b4mad.net 	&	291	&	1.00	&	393 \\
		thefigtrees.net & 	89	&	1.00	&	102	&	derby.ac.uk	&	137	&	1.00	&	197 \\
		bblfish.net & 	82	&	0.99	&	150	&	heppnetz.de	&	121	&	1.00	&	199 \\
		wikier.org & 	96	&	1.00	&	133	&	ivan-herman.net	&	196	&	1.00	&	303 \\
		bl.uk & 	102	&	0.46	&	246	&	jones.dk	&	164	&	1.00	&	155 \\
		kanzaki.com & 	176	&	0.99	&	294	&	ldodds.com	&	115	&	1.00	&	125 \\
		taxonconcept.org & 	139	&	0.92	&	424	&	lmco.com	&	128	&	1.00	&	204 \\
		fundacionctic.org & 	110	&	0.97	&	390	&	mfd-consult.dk	&	192	&	1.00	&	315 \\
		data.gov.uk & 	258	&	0.93	&	920	&	mit.edu	&	174	&	0.96	&	293 \\
		bbc.co.uk & 	146	&	1.00	&	522	&	nickshanks.com	&	100	&	0.97	&	164 \\
		\hline
	\end{tabular}
\end{table*}
To validate TermPicker's recommendation quality, we perform two separate evaluations.
One evaluation uses the seed-list data of the Dynamic Linked Data Observatory (DyLDO)~\cite{dyldo}\footnote{\url{http://swse.deri.org/dyldo/}, last access 12/12/15} and the other evaluation uses the Billion Triple Challenge dataset (BTC) \linebreak 2014~\cite{btc2014}\footnote{\url{http://km.aifb.kit.edu/projects/btc-2014/}, last access 12/12/15} (crawl no. $1$).
We chose these two data sets, as they represents parts of the LOD cloud in different way.
For once, DyLDO's seed list, i.e., the set of URIs that form the core of the data crawling, is different from the seed list of the BTC 2014 dataset.
The seed list of the BTC 2014 dataset is sampled from the previous year's dataset, and the initial seed-list was gathered from various semantic search engines.
DyLDO's seed list comprises the $220$ most popular URIs selected from the BTC 2011\footnote{\url{https://km.aifb.kit.edu/projects/btc-2011/}, last access 12/12/15} dataset based on a PageRank analysis combined with another $220$ URIs from the CKAN/LOD\footnote{\url{https://datahub.io/dataset?tags=lod}, last access 12/12/15} registry, which were not contained in the BTC 2011 dataset.
This means that the \dyldo and the BTC 2014 datasets contain different data, as it was crawled from different dataset on the LOD cloud.
Furthermore, DyLDO's seed list makes about $50\%$ of the entire data contained in the \dyldo dataset, whereas the seed list of BTC 2014 makes only less than one percent, resulting in way more data than in \dyldo.

DyLDO comprises a considerable amount of about $10.8$ million triples from $382$ different pay-level domains.
In total there are about $2.3$ million distinct vocabulary terms from about $600$ vocabularies.
The BTC 2014 dataset contains about $1.4$ billion triples, of which we use solely $34$ millions, in order to keep the in-memory SLP computation maintainable.
These $34$ million triples are provided by $3,493$ pay-level domains.
Within these triples there are about $5.5$ million distinct RDF types and properties from about $1,530$ different vocabularies.

We regard a vocabulary simply by its URI namespace which is specified by the W3C.\footnote{\url{http://www.w3.org/2001/sw/BestPractices/VM/http-examples/2006-01-18/#naming}, last access 9/25/15}
This means that a vocabulary either uses a \emph{hash namespace} or a \emph{slash namespace}, i.e., the vocabulary of a term is represented by the URI before the last occurrence of either a hash or a slash.
Therefore, we do not distinguish between a vocabulary being of type \code{owl:Ontology}, \code{voaf:Vocabulary}, or others, and keep it as simple as possible.
To differentiate which dataset is under the control of which data publisher, we make use of the the pay-level domain (PLD) calculated from the the context-URI $c$ contained in the data.
A pay-level domain (PLD) is a direct sub-domain of a top-level domain, such as \emph{.org} or \emph{.com}, or of a second-level country domain, such as \emph{.de} or \emph{.uk}.\footnote{To calculate the pay-level domain, we make use of the Google guava library: \url{https://code.google.com/p/guava-libraries/}, last access 9/28/15} 
Examples of pay-level domains included in the BTC 2014 (about $3,500$ PLDs) and the \dyldo (about $382$ PLDs) dataset are \code{dbpedia.org} or \code{bbc.co.uk}.

A fully qualified domain name, such as the context-URI itself, would over-exaggerate the diversity of the data, as it would also differentiate data from different sub-domains.
Hence, by referring to a \emph{dataset published on the LOD cloud} or \emph{a data publisher on the LOD cloud}, we refer to a PLD that specifies which data publisher is in control of the data.

For each evaluation, the evaluation dataset is split by the pay-level domains.
The data from ten different PLDs is used as training and test set, whereas the data from the remaining PLDs is used to simulate the data sets published on the LOD cloud. 
For each fold of the 10-fold leave-one-out evaluation, one of the ten PLDs is left out and resembles the test set, whereas the other nine PLD represent the training set.
As mentioned before, both the test and training set consists of the computed SLPs from the data of the according pay-level domain(s).

The more query-SLPs are used to train and test the ranking model, and the larger the data for calculating the features values, the more representative are the generated results~\cite{busa2012apple}.
Thus, the ten pay-level domains for training and testing are selected based on two criteria.
\begin{enumerate}[label=($C$\arabic*),leftmargin=0.8cm,itemsep=1ex]
	\item A high number of distinct vocabulary terms within a PLD 
	\item A high ratio between the reused vocabulary terms and all RDF types and properties used within a PLD
\end{enumerate}
The high number of distinct vocabulary terms indicates that resources of various RDF types are interlinked via several different properties.
This way, it is very likely to calculate a high number of distinct SLPs from that data.
A negative example is a dataset modeling several million instances of type \code{foaf:Person} knowing other persons, as this will generate solely one SLP.
The high ratio between the reused terms and all terms used to describe the data indicates that most resources and their interlinking are described via reused and not self-defined vocabularies.
This enables to calculate SLPs that most likely contain many reused terms, which is important to generate valuable recommendations.
Selecting PLDs for training and test sets randomly and not based on $C1$ and $C2$ is very likely to result in poor evaluation results, as many PLDs either do not use many different vocabulary terms or they use many self-defined terms.

Table~\ref{tab:evalDataStats} provides an overview of the selected PLDs used for the evaluations based on the \dyldo (left half of the table) and BTC 2014 (right half of the table) dataset as well as the numbers considering $(C1)$ and $(C2)$.
Those PLDs that provided the highest numbers in both $C1$ and $C2$ were selected as test and training sets.
Furthermore, it displays the number of distinct SLPs that are calculated from the data of the selected pay-level domains.

Naturally, SLPs that are used to train the ranking model are different to the SLPs that are used to evaluate the model.
The data from the remaining PLDs that is used for calculating the features contains $117,776$ (\dyldo) and $227,010$ (BTC 2014) SLPs, respectively.

\subsection{Evaluation Metrics}\label{sec:metrics}
As a user, who searching for possible RDF types and properties for reuse, is likely to browse only through the top-$k$ vocabulary terms (where $k$ is generally a small number such as $5$ or $10$), it is important to evaluate the ranking model by measures that use ordered sets of vocabulary terms.
We use the Mean Average Precision (MAP) and the Mean Reciprocal Rank to the fifth position (MRR@5). 
Both measures illustrate the quality of the ranking model well, as they compute values using such ordered sets of vocabulary terms (in contrast to basic measures such as precision and recall).

On one hand, MAP provides a measure of quality across recall levels~\cite{manning2008introduction}.
It illustrates the quality of the entire result list in which the ranking position of the relevant vocabulary term is considered.
The higher the MAP value, the more relevant vocabulary terms are ranked to the top positions of the result list.
On the other hand, the Mean Reciprocal Rank at the first $k$ results (MRR@k) investigates the result list only to the rank position of the first relevant vocabulary term~\cite{reciprocalRank}.
In other words, MRR returns a metric specifying the ranking position of the first relevant term.

In the following, we use $k=5$.
We define the set of query-SLPs as $Q = \{slp_{q_1},...,slp_{q_n}\}$.
If the set of relevant vocabulary terms for a query $slp_{q_j} \in Q$ is $\{rt_1, \ldots, rt_{m_j}\}$ and $R_{jh}$ ($1 \leq h \leq m_j$) is the set of ranked retrieval results from the top result until one gets to the relevant vocabulary term $rt_h$,  
then the \textbf{Mean Average Precision} and the \textbf{Mean Reciprocal Rank} of $Q$  defined as
\begin{equation}
\begin{aligned}
	\mbox{MAP}(Q) &= \frac{1}{|Q|} \sum_{j=1}^{|Q|} \frac{1}{m_j} \sum_{h=1}^{m_j} \mbox{Precision}(R_{jh})
\end{aligned}
\end{equation}
\begin{equation}
\begin{aligned}
	\mbox{MRR}(Q) &= \frac{1}{|Q|} \sum_{j=1}^{|Q|} \frac{1}{|R_{jh}|}
\end{aligned}
\end{equation}

\section{Results}\label{results}
\begin{figure*}[t]
\centering
\includegraphics[width=1\textwidth]{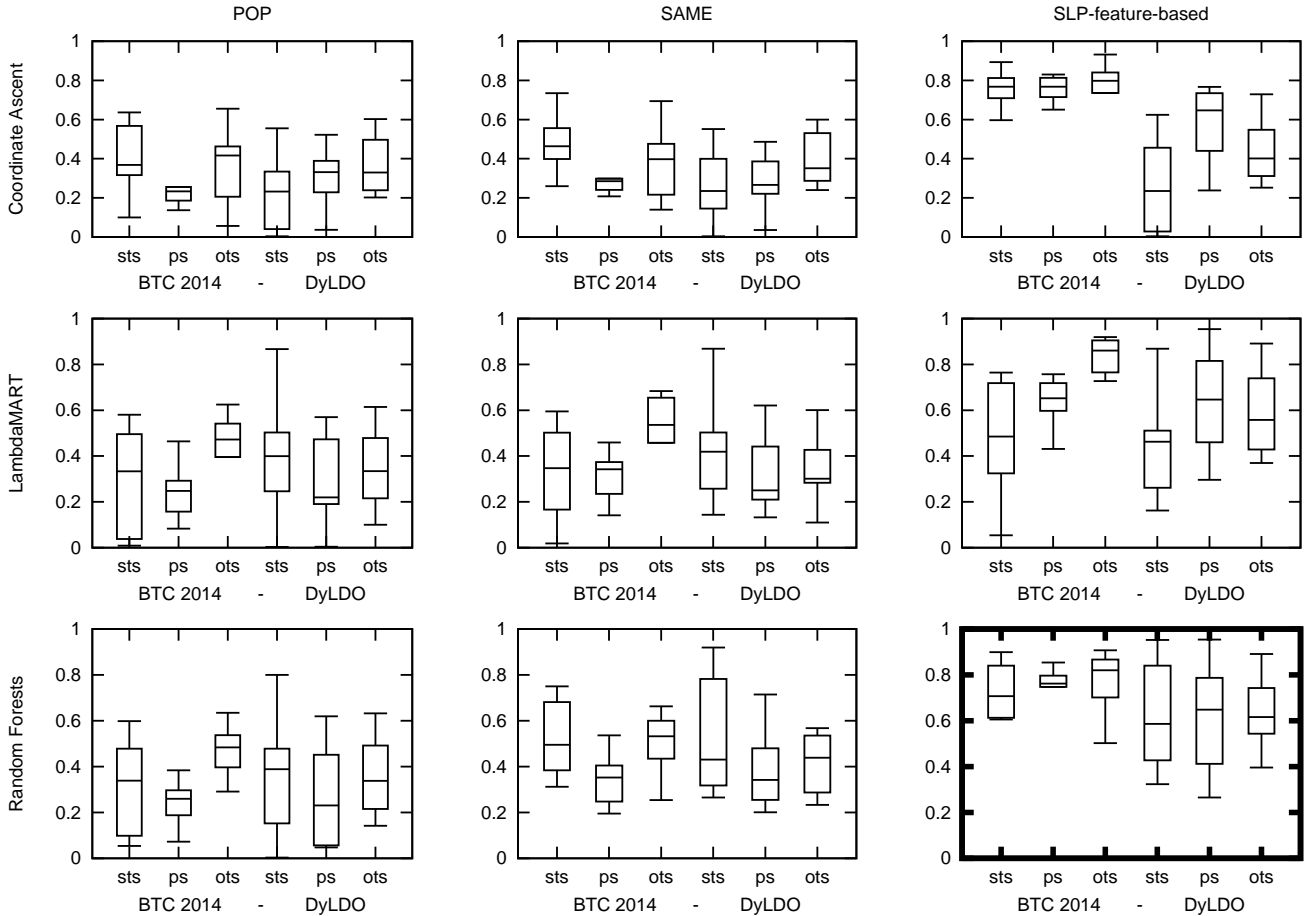}
\caption{\textbf{MAP results.}
		On the x-axis of each plot one finds the recommendations for RDF types for resources in subject position ``sts'', for properties ``ps'', of for RDF types for resources in object position ``ots''.
		The left part of each plot represents the results of the evaluation performed on the BTC 2014 dataset and the right part of the plots depicts the results using the \dyldo dataset.
		The proposed SLP-Feature can be compared with the baseline reusing popular vocabularies (POP) and the baseline reusing popular vocabularies from the same vocabulary (SAME) for the three most competitive L2R algorithm from the RankLib library.
		The plot marked bold depicts the overall best results, which is the Random Forests algorithm using the SLP-Feature.}
\label{fig:results_a}
\end{figure*}
\begin{figure*}[t]
\centering
\includegraphics[width=1\textwidth]{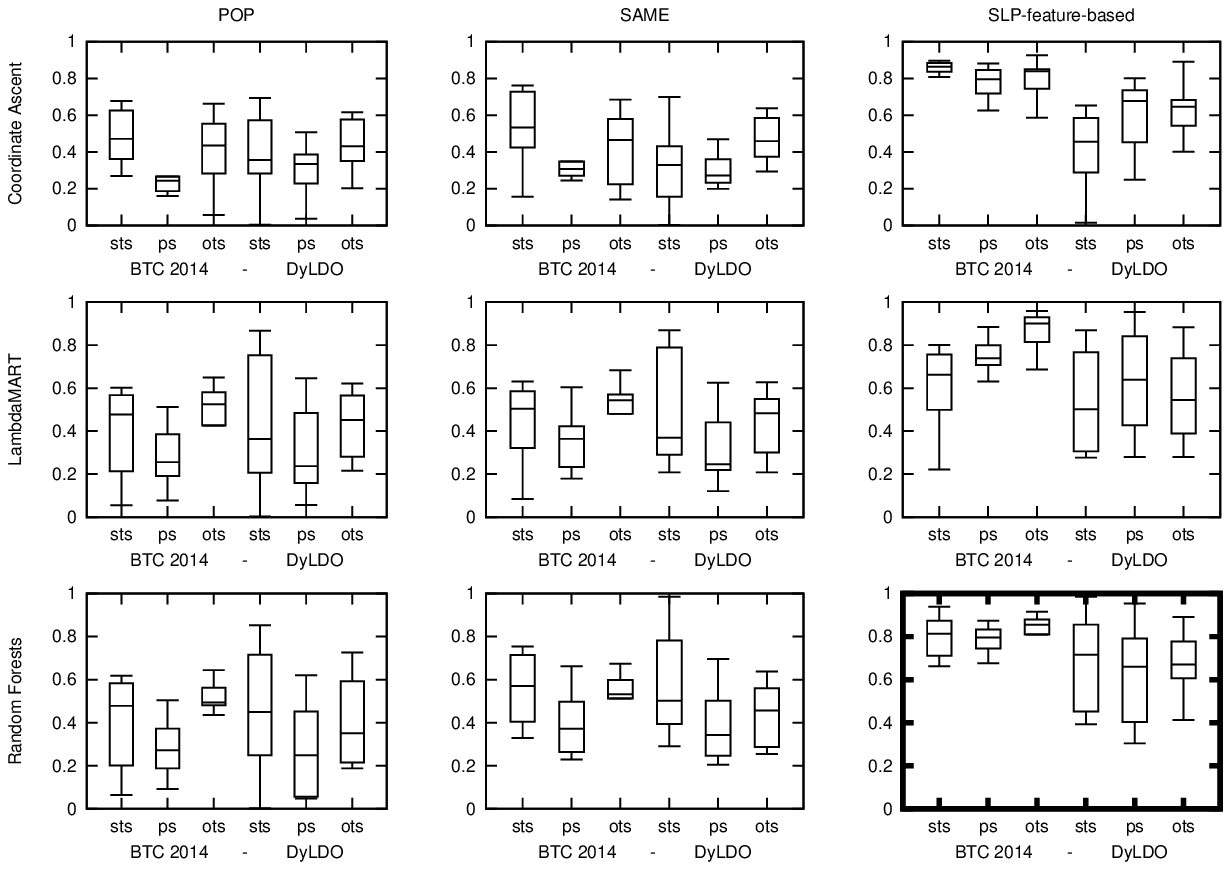}
\caption{\textbf{MRR@5 results.}
		On the x-axis of each plot one finds the recommendations for RDF types for resources in subject position ``sts'', for properties ``ps'', of for RDF types for resources in object position ``ots''.
		The left part of each plot represents the results of the evaluation performed on the BTC 2014 dataset and the right part of the plots depicts the results using the \dyldo dataset.
	The proposed SLP-Feature can be compared with the baseline reusing popular vocabularies (POP) and the baseline reusing popular vocabularies from the same vocabulary (SAME) for the three most competitive L2R algorithm from the RankLib library.
		The plot marked bold depicts the overall best results, which is the Random Forests algorithm using the SLP-Feature.}
\label{fig:results_b}
\end{figure*}
The results of the evaluation are presented in Figure~\ref{fig:results_a} and Figure~\ref{fig:results_b}.
They illustrate the recommendation quality via box-plots based on the MAP and the MRR@5 respectively.
The figures depict the measurements of the recommendation quality considering the aspects (i), (ii), and (iii) introduced in Section~\ref{sec:eval}.
The three most competitive L2R algorithms in the RankLib library are: \emph{Coordinate Ascent}, \emph{LambdaMART} and the \emph{Random Forest} algorithm.
The difference between these three L2R algorithms can be observed by comparing the three different rows in Figures~\ref{fig:results_a} and~\ref{fig:results_b}.
The varying recommendation quality between the different set of features can be examined by comparing the three columns of the Figures.
Both reusing solely popular vocabulary terms (marked as POP) and reusing vocabulary terms from the same vocabulary (marked as SAME) resemble the baseline, as they are considered current state of the art strategies to reuse a vocabulary~\cite{2014SchaibleESWC,series/synthesis/2011Heath}.
Our proposed approach, marked as ``SLP-feature-based'', additionally uses the SLP-feature. 
Within each plot, the x-axis displays the different recommendations of a RDF type for resources in subject position (abbreviated as ``sts''), of a RDF type for resources in object position (abbreviated as ``ots''), or of a property (abbreviated as ``ps'') for both the BTC 2014 and the \dyldo dataset.
Each box plot comprises the measured recommendation quality of the ten PLDs that were used as test sets in the 10-fold leave-one-out evaluation.
The plot that is marked bold illustrates the configuration, i.e., which features and which L2R algorithm, achieving the overall best recommendation quality.

\begin{table*}[t]
\begin{center}
	\caption{\textbf{MAP and MRR@5 values for BTC 2014.} Each row depicts the average MAP and MRR@5 values and their standard deviation for the three most competitive L2R algorithms in the RankLib library and the set of features, i.e., baseline POP, baseline SAME, and using the SLP-feature. 
	The columns depict the difference between recommending a property or RDF types for resources at subject or object positions of a triple.
	The overall recommendation quality of a L2R algorithm with a specific set of features is illustrated in the two most right columns}
	\label{tab:averageNumsBTC}
	\begin{tabular}{r  l || r r r r r r | r r }
& &\multicolumn{2}{c}{sts}&\multicolumn{2}{c}{ps}&\multicolumn{2}{c}{ots}&\multicolumn{2}{c}{overall}\\
Model & Featues & MAP & MRR@5 & MAP & MRR@5 & MAP & MRR@5 & MAP & MRR@5 \\
\hline\hline
CoordinateAscent&POP		&.38 (.18)&.49 (.15)&.25 (.11)&.27 (.11)&.38 (.19)&.41 (.19)&.34 (.16)&.39 (.15)\\
							&SAME	&.48 (.16)&.55 (.19)&.31 (.10)&.33 (.09)&.39 (.18)&.43 (.18)&.39 (.15)&.44 (.15)\\
							&SLP		&.75 (.12)&.83 (.10)&.76 (.06)&.78 (.08)&.76 (.14)&.81 (.10)&.76 (.11)&.81 (.09)\\
LambdaMART		&POP		&.31 (.22)&.39 (.21)&.27 (.15)&.28 (.14)&.42 (.20)&.45 (.20)&.33 (.19)&.37 (.18)\\
							&SAME	&.34 (.21)&.44 (.17)&.33 (.13)&.34 (.14)&.49 (.19)&.49 (.17)&.39 (.18)&.42 (.16)\\
							&SLP		&.46 (.25)&.61 (.18)&.64 (.10)&.73 (.12)&.82 (.12)&.86 (.09)&.64 (.16)&.73 (.13)\\
RandomForests	&POP		&.32 (.20)&.40 (.21)&.26 (.12)&.28 (.12)&.45 (.17)&.48 (.15)&.34 (.16)&.39 (.16)\\
							&SAME	&.52 (.16)&.56 (.15)&.37 (.14)&.39 (.14)&.49 (.16)&.50 (.17)&.46 (.15)&.48 (.15)\\
							&SLP		&.72 (.11)&.80 (.10)&.75 (.10)&.77 (.10)&.78 (.12)&.83 (.08)&.75 (.11)&.8 (.09)
\end{tabular}
\end{center}
\end{table*}

\begin{table*}[t]
\begin{center}
	\caption{\textbf{MAP and MRR@5 values for \dyldo.} Each row depicts the average MAP and MRR@5 values and their standard deviation for the three most competitive L2R algorithms in the RankLib library and the set of features, i.e., baseline POP, baseline SAME, and using the SLP-feature. 
	The columns depict the difference between recommending a property or RDF types for resources at subject or object positions of a triple.
	The overall recommendation quality of a L2R algorithm with a specific set of features is illustrated in the two most right columns}
	\label{tab:averageNumsDyLDO}
	\begin{tabular}{r  l || r r r r r r | r r }
	& &\multicolumn{2}{c}{sts}&\multicolumn{2}{c}{ps}&\multicolumn{2}{c}{ots}&\multicolumn{2}{c}{overall}\\
Model & Featues & MAP & MRR@5 & MAP & MRR@5 & MAP & MRR@5 & MAP & MRR@5 \\
\hline\hline
CoordinateAscent&POP		&.22 (.18)&.37 (.23)&.31 (.14)&.31 (.14)&.37 (.15)&.43 (.13)&.30 (.16)&.37 (.17)\\
							&SAME	&.26 (.16)&.33 (.24)&.29 (.13)&.29 (.13)&.39 (.13)&.47 (.12)&.31 (.14)&.36 (.16)\\
							&SLP		&.25 (.23)&.43 (.21)&.58 (.18)&.60 (.19)&.45 (.17)&.63 (.14)&.43 (.19)&.55 (.18)\\
LambdaMART		&POP		&.48 (.27)&.54 (.33)&.38 (.28)&.39 (.27)&.43 (.24)&.48 (.16)&.43 (.26)&.47 (.25)\\
							&SAME	&.48 (.26)&.57 (.29)&.40 (.26)&.40 (.26)&.41 (.23)&.51 (.19)&.43 (.25)&.49 (.25)\\
							&SLP		&.49 (.27)&.56 (.27)&.63 (.23)&.63 (.24)&.58 (.20)&.56 (.21)&.57 (.23)&.58 (.24)\\
RandomForests	&POP		&.44 (.29)&.55 (.31)&.35 (.28)&.36 (.28)&.43 (.25)&.49 (.26)&.41 (.27)&.47 (.28)\\
							&SAME	&.59 (.27)&.65 (.24)&.46 (.24)&.46 (.24)&.49 (.21)&.52 (.21)&.51 (.24)&.54 (.23)\\
							&SLP		&.65 (.26)&.70 (.24)&.63 (.25)&.63 (.24)&.64 (.17)&.68 (.15)&.64 (.23)&.67 (.21)
\end{tabular}
\end{center}
\end{table*}

\paragraph{(i)~Differences between L2R algorithms}
Comparing the three most competitive L2R algorithms, one can observe that there are no obvious differences between the algorithms when using solely features $f_1 - f_3$ (baseline POP) or when using features $f_1 - f_4$ (baseline SAME).
The median MAP and MRR@5 values are between $0.3$ and $0.5$ for each of the three algorithms.
However, when making use of all features including the SLP-feature, the differences of the median values are more noticeably.
While the median values using the algorithms \emph{Coordinate Ascent} and \emph{Random Forests} on the BTC 2014 data are between $0.7$ and $0.8$, the median values using \emph{LamdaMART} vary  in average at $0.6$.
Four other algorithms from the RankLib library, i.e., \emph{AdaRank}\cite{xu2007adarank}, \emph{RankNet}\cite{burges2005learning}, \emph{RankBoost}~\cite{rankBoost}, and \emph{ListNet}~\cite{listNet}, did not provide such good results.
The median MAP and MRR@5 values were never above $0.3$, and there was no increase of the recommendation quality between using the different sets of features.
Finally, the L2R algorithm \emph{MART}~\cite{burges2005learning} was able to achieve a median MAP and MRR@5 value of about $0.5$, but in total, MART's successor, i.e., LambdaMART, provided very similar but slightly better results.
These results can be observed when using the BTC 2014 dataset as well as the \dyldo dataset as evaluation data.

\paragraph{(ii)~Impact of the SLP-feature}
Comparing the different set of utilized features, one can observe that the differences are more visible when using the BTC 2014 dataset as evaluation data.
There is a slight increase in the recommendation quality, when adding feature $f_4$ to the set of features, i.e., the medians for the baseline POP and the baseline SAME differ in average by about $7\%$.
When adding the SLP-feature however, the median recommendation quality increases by about $30\%$ compared to the baseline of reusing solely popular vocabulary terms (compared to POP).
Even compared to the SAME baseline, i.e., reusing popular vocabulary terms from the same vocabulary, one can perceive an increase of the recommendation quality by $20\%$.
Such differences between the sets of utilized features are not as visible when performing the evaluation on the \dyldo dataset.
However, one can still observe that there is only a small increase of the recommendation quality ($< 7\%$) of the baseline SAME compared to the baseline POP.
Using also the SLP-feature increases the median recommendation quality by about $15 - 20\%$ compared to the baselines POP and SAME.

\paragraph{(iii)~Differences between recommendation types}
Finally, using all features (including the SLP-feature) and comparing the recommendation quality between recommending RDF types for resources in subject position, RDF types in object position, or properties, only slight changes (between $5-10\%$) in the recommendation quality can perceived.
Solely the L2R algorithm \emph{LambdaMART} based on the BTC 2014 dataset has a higher median recommendation quality when suggesting RDF types for resources in object position ($\text{MAP} = .83$) compared to the medians when suggesting a property ($\text{MAP} = .63$) and when suggesting an RDF type for a resource in subject position ($\text{MAP} = .5$).
The MRR@5 values are very much the same.

In addition, Table~\ref{tab:averageNumsBTC} and Table~\ref{tab:averageNumsDyLDO} illustrate the average MAP and MRR@5 values (including the standard deviation) for the evaluations based on the BTC 2014 and the \dyldo datasets, respectively.
They underline the increase of the recommendation quality, when adding the SLP-feature to the set of features, which is used by the ranking model.
For the BTC 2014 dataset, in average, using the SLP-feature provides a higher MAP and MRR@5 value than using features to reusing terms from the same vocabulary (SAME) by $29\%$, and comparing to the features for reusing solely popular vocabulary term (POP), it provides better recommendations by $36\%$.
For the \dyldo data, these differences are not as distinctive, but they are still $13\%$ compared to the baseline SAME and $23\%$ compared to the baseline POP.
Looking at Table~\ref{tab:averageNumsBTC}, the L2R algorithm \emph{Coordinate Ascent} seems to provide the best results with a MAP of $\text{MAP} =.76$ and an MRR@5 value of $\text{MRR@5} = .81$.
However, it does not perform as well based on the \dyldo dataset ($\text{MAP} = .43$ and $\text{MRR@5} = .55$).
Therefore, the overall best recommendation quality, which is calculated based on the values from Table~\ref{tab:averageNumsBTC} and Table~\ref{tab:averageNumsDyLDO}, is provided by the L2R algorithm \emph{Random Forests} using all features, including the SLP feature ($\text{MAP} = .70$ and $\text{MRR@5} = .73$).

\section{Discussion}\label{discussion}
The discussion is structured as follows:
In Section~\ref{sec:discResults}, we discuss the results of the evaluation based on the three main contributions of this paper, i.e., 
(i)~the difference between the utilized Learning To Rank algorithms,
(ii)~the impact of the SLP-feature on the recommendation quality, and
(iii)~the difference between recommending RDF types and properties.
We also provide insights whether the measured differences are significant using the Friedman test (differences are significant with a $p$-value $p < .05$) and a Wilcoxon signed-rank test with a Bonferroni correction applied to detect pair-wise differences (the corrected $p$-value for (i) to (iii) is $p < (.05 / 3 = .017)$). 
In Section~\ref{sec:discApproach}, we discuss the general use of a Learning to Rank algorithm for providing vocabulary term recommendations, as well as the limitations of the utilized evaluation design.

\subsection{Discussion of the Results}\label{sec:discResults}
\subsubsection{Differences between the L2R algorithms~(i)}
From the eight L2R algorithms contained in the RankLib library, solely four algorithms were able to provide recommendations with an MAP above $50\%$ when making use of all features.
Out of the four algorithms with  $\text{MAP} < 0.5$, two algorithms (\emph{RankNet} and \emph{RankBoost}) are pair-wise approaches, and the other two algorithms (\emph{ListNet} and \emph{AdaRank}) are list-wise approaches.
The best performing algorithm, i.e., \emph{Random Forests}, is a point-wise approach, whereas the other ones (\emph{Coordinate Ascent}, \emph{LambdaMART}, and \emph{MART}) are all list-wise approaches. 

Generally, list-wise and pair-wise approaches perform better than point-wise approaches~\cite{l2rCompare, busa2012apple}.
However, in cases where there is only a binary relevance, i.e., a recommendation candidate is either relevant or irrelevant, point-wise approaches perform better, if there is solely one relevant recommendation candidate for most queries ~\cite{l2rCompare, busa2012apple}.
In our use-case, recommendation candidates have indeed a binary relevance.
Additionally, most of the query-SLPs used to train and evaluate the ranking model contained mostly up to three vocabulary terms.
Therefore, based on the evaluation design, only one or two vocabulary terms could be extracted, to provide relevant recommendation candidates and to provide TermPicker with a non-empty query-SLP.
Thus, in our evaluation, we use a binary relevance, and for most of the queries there are solely one or two relevant recommendation candidates.
Based on this, it is quite reasonable that a point-wise L2R algorithm performs best.

This is underlined by the significant differences between the recommendation quality using the algorithm \emph{Random Forests} and the recommendation quality using the other L2R algorithms.
The Friedman test, which compares the overall MAP and MRR@5 values based on both the BTC 2014 and the \dyldo data using all features, showed that these differences are statistically significant with $\mathcal{X}^2 = 14,000, p = .001$.
The Wilcoxon signed-rank with Bonferroni correction applied proved that there is no significant difference between using the \emph{Coordinate Ascent} and \emph{LambdaMART} algorithm ($Z = -0.243$, $p = .808$  \textbf{n.s.}).
However, with $Z = -2.492, p = .013$, the \emph{Random Forests} algorithm provides significantly better recommendation than the \emph{Coordinate Ascent} algorithm, and with $Z = -4.237, p < 0.001$ it is also significantly better than \emph{LambdaMART}.

\subsubsection{Impact of the SLP-feature~(ii)}
To discuss the impact of the features on the recommendation quality, we use the best performing L2R algorithm for each set of features across both the BTC 2014 and the \dyldo dataset, i.e., for the baseline POP that is the L2R algorithm \emph{LambdaMART} and for the baseline SAME as well as for using the SLP-feature that is the algorithm \emph{Random Forests}.

With $\text{MAP} \approx .35$, the average MAP value of recommendations based on reusing solely popular vocabulary terms (baseline POP) is quite high.
Specifically considering the fact, that the feature values describing the popularity of a recommendation candidate are static, meaning they do not depend on the query-SLP.
However, such MAP values can be explained by the setup of our evaluation.
As we use real-life data for our evaluation, the relevant recommendation candidates are vocabulary terms that actually have been used by some ontology engineer to describe the data.
The best practices~\cite{series/synthesis/2011Heath} recommend to reuse terms from popular vocabularies, therefore it is very likely that the ontology engineer initially has reused terms from popular vocabularies.
This leads to a general coherence, which is trained by the Learning To Rank algorithm, that a vocabulary term from a popular vocabulary is likely to be a relevant recommendation candidate.

Recommendations based on reusing vocabulary terms from the same vocabulary (baseline SAME) have an MAP value of $\text{MAP} \approx .43$.
A Friedman test ($\mathcal{X}^2 = 51,667, p < .001$) and the following Wilcoxon signed-rank test ($Z = -1.692, p = .011$) indicate that this difference in the recommendation quality is still significant compared to the baseline POP.
However, it seems interesting that using the same-vocabulary-feature provides only an $8\%$ gain in the absolute recommendation quality.
Investigating the vocabulary terms used in the query-SLPs showed that many query-SLPs contain quite popular vocabulary terms, but they are rarely from the same vocabulary.
In total, in $43\%$ of the SLPs in the training and test set contained two or more terms from the same vocabulary.
That means: the vocabulary terms that are extracted from an SLP before providing TermPicker with the resulting query-SLP are rarely from the same vocabulary as the remaining terms in the query-SLP. 
Thus, the L2R algorithms are less likely to regard this feature to provide more appropriate recommendations.

Using the SLP-feature increases the average MAP value up to \linebreak $\text{MAP} \approx .70$.
A Wilcoxon signed-rank test showed that using the SLP-feature and comparing its recommendation quality to the one of the baseline SAME, the $p$-value is $Z = -4.782, p < 0.001$.
Due to the transitivity of this relation, the recommendation quality when using the SLP-feature is also significantly higher to the recommendation quality when using solely features to define popular vocabulary terms (baseline POP).
Such a result depict to which extend the SLP-feature is relevant for providing valuable vocabulary term recommendations.
Yet again, these results are based on using real-life data for calculating the query-SLPs for the evaluation.
If the recommendation quality using the SLP-feature is that large, one can argue that the utilized real-life data was initially modeled by investigating which vocabulary terms other data providers have used to model their data.
However, as establishing an ontological agreement in data representation is one central goal when reusing vocabularies~\cite{series/synthesis/2011Heath}, the results indicate that using recommendation based on the SLP-feature will eventually result in such a goal.

The evaluation based on the BTC 2014 data provides a more noticeable gain in the recommendation quality when using the SLP-feature than the evaluation based on the \dyldo data.
In general, the key aspect of providing valuable recommendations lies in training the ranking model using representative data.
In our case, this includes the query-SLPs that are used to train the ranking model, but also the data that is used to calculate the five feature values for each recommendation candidate.
Further investigations have shown, that the feature values calculated based on \dyldo data were less expressive compared to the feature values calculated based on the BTC 2014 dataset.
In other words, the evaluation based on the BTC 2014 dataset provided an SLP-feature value of $f_5 > 0$ for $37\%$ more relevant recommendation candidates than using the \dyldo data.
The ranking model, which was learned based on the BTC 2014 data, therefore ranked recommendations with an SLP-feature value greater than zero rather to the top of the result list.
This observation is validated by using the ranking models learned using the BTC 2014 data to rank the recommendation candidates for query-SLPs calculated from the \dyldo data.
The resulting recommendation quality was $15\% - 20\%$ higher than using a ranking model learned based on the \dyldo data.
The reason for such a difference in the recommendation quality is very likely the number of SLPs in the set $\mathbb{SLP}$, i.e., the SLPs that are calculated from existing datasets on the LOD cloud.
Using the BTC 2014 dataset the number of such SLPs is twice as high compared to the number of such SLPs using the \dyldo data.
As it is much more likely to calculate an SLP-feature value of $f_5 > 0$ with more SLPs contained in $\mathbb{SLP}$, it is quite reasonable that the evaluation based on the BTC 2014 data provides a higher recommendation quality.

\subsubsection{RDF type recommendations vs. property recommendations~(iii)}
The differences between recommending RDF types a properties represent the different \emph{modeling steps} in the engineering process of a schema~\cite{Noy101}.
It is accustomed to define a set of classes, which depict the entities that one wants to model, first, and then define relationships connecting these classes.
Thus, it could be also accustomed that TermPicker recommends RDF types to describe the defined classes before recommending properties to interlink the RDF types.
However, the differences in the recommendation quality between recommending RDF types for resources in subject or object position, or recommending properties seem to be marginal and cannot be considered significant according to the Friedman test, $\mathcal{X}^2 = 14,000, p = .449$  \textbf{n.s.}.

One aspect might be that the recommendation quality depends on how many vocabulary terms are already included in the query-SLP.
In other words, a query-SLP containing three or more vocabulary terms could provide more concrete recommendations, than a query-SLP containing solely one term.
For example, one would assume, that on one hand the query SLP $slp_{q_1}$ with
\begin{equation*}
	\begin{aligned}
		slp_{q_1} = (\{\code{foaf:Person}\}, \varnothing, \{\code{foaf:Image}\})
	\end{aligned}
\end{equation*}
produces more specific recommendation, due to the restriction of already reusing \code{foaf:Person} and \code{foaf:Image}.
On the other hand a query-SLP $slp_{q_2}$, such as
\begin{equation*}
	\begin{aligned}
		slp_{q_2} = (\{\code{foaf:Person}\}, \varnothing, \varnothing)
	\end{aligned}
\end{equation*}
should produce a larger amount of recommendations, as the query is not as restricted as the query $slp_{q_1}$.
The chances of ranking a relevant vocabulary term to the top of the result should thus be higher for a query-SLP such as $slp_{q_1}$, i.e., query-SLPs that contain more vocabulary terms, as there is not as much noise in the recommendations.
However, the differences between the query-SLPs with varying amount of contained vocabulary terms did not prove to be significant, $\mathcal{X}^2 = 15,800, p = .327$  \textbf{n.s.}.
Therefore, one can conclude that TermPicker provides appropriate vocabulary term recommendations regardless if one is searching for RDF types describing resources in subject or object position of a triple, or for properties connecting two sets of RDF types.
If another dataset on the LOD cloud uses a vocabulary term in conjunction with the terms included in the query-SLP, it has a large chance to be ranked at the top of the recommendation list.

\subsection{Discussion of the Proposed Approach and the Evaluation}\label{sec:discApproach}
Learning to Rank tries to establish a correlation between the feature values of a recommendation candidate and its relevance~\cite{liu2009learning}.
Using the SLP-feature provides valuable results in most cases, but in the end the ranked results lists depend on the ranking model.
Whether or not the SLP-feature is useful thus depends on the utilized training data, as demonstrated by the differences of using the BTC 2014 and the \dyldo data.
For \dyldo, it does not work as well and leads to a decrease of the influence of the SLP-feature.
This is because it does not contain a large variety of vocabulary terms and thereby decreases the chance of finding a term that has been used by other datasets on the LOD cloud in a similar way.
The same applies for the same-vocabulary-feature.
Generally, the proposed recommendation approach is reproducible with each Linked Data collection, e.g., with the BTC 2012 or the \emph{Timbl} dataset which seed list contains URIs from Tim Berners-Lee's FOAF profile, but the bigger the data, the better the training data and the resulting ranking model.
The best option would be to use the data from all datasets on the LOD cloud.
However, computing SLPs from such a massive data collection is very time consuming and was not feasible for the provided evaluation.

The problem of finding an appropriate vocabulary term is a typical information retrieval problem that can be addressed via a machine learning approach.
Thus, we validated the usefulness of the SLP-feature by using Learning To Rank, as it provides a methodology to induce a ranking model, that can be applied in general situations to retrieve appropriate vocabulary terms for reuse.
Other approaches such as the Data Mining approach \emph{Association Rules} conquer this problem by recommending terms based on the simple statement: ``Datasets on the LOD cloud, who have used these vocabulary terms, have also used the following:...''.
This way, a vocabulary term that is not used in a similar manner will not be recommended.
However, it also increases the chances that the result lists return empty.
Therefore, it is rather a question whether the user also wants to get recommendation that make him/her ``think outside the box'', or whether he/she likes to stay as conform as possible to what others have used.

A potential threat to the validity of our experiments is the utilized evaluation design. 
It considers solely the recommendation candidates as relevant that have been extracted from a query-SLP before providing this query-SLP as input for TermPicker (cf. Section~\ref{sec:experimentDesign}).
This leads to two major vulnerabilities considering the validity of the evaluation.
For once, many recommendation candidates are identified as irrelevant, although they are appropriate considering the \code{rdfs:domain} and \code{rdfs:range}, the \code{owl:equivalentClass}, or other information. 
For example, for the query-SLP $slp_q$ with
\begin{equation*}
\begin{aligned}
	slp_j 	&= (\{\code{swrc:Publication}\}, \{\code{swrc:author}\}, \{\code{foaf:Person}\}) \\
	slp_q 	&= slp_j \ominus_{ps} \code{swrc:author} \\
				&= (\{\code{swrc:Publication}\},  \varnothing, \{\code{foaf:Person}\})
\end{aligned}
\end{equation*}
the only relevant recommendation candidate for properties is \code{swrc:}\code{author}, as it was originally used.
Properties, such as \code{dc:creator} or \code{foaf:maker} are considered as irrelevant in our evaluation, although it would make sense to reuse these properties to interlink resources of type \code{swrc:Publication} with resources of type \code{foaf:Person}.
Thus, an L2R algorithm may identify many \emph{appropriate} vocabulary terms (with an SLP-feature greater than zero) as irrelevant, which then can result in a ill-trained ranking model.
Second, using many SLPs such as $slp_j$ in the previous example, will favor point-wise L2R algorithms, as they tend to perform better, if there is only one or a few relevant items~\cite{liu2009learning}.
The previous example also shows, that there might be more than only a few relevant vocabulary terms. 
Utilizing a bigger set of relevant recommendation candidates might change the quality of point-wise, pair-wise, and list-wise L2R algorithms, such that list-wise and pair-wise algorithms might perform better than the point-wise approach.
However, addressing this limitation requires human judgment whether a recommendation is relevant or not.
Thus, conducting an experiment with human users is part of our future work.

\section{Related Work}\label{relWork}
The related work focuses on the schema-level patterns as well as on services that support an engineer in reusing vocabularies.
The notion of schema-level pattern can be compared to the notion of so-called \emph{triple patterns}~\cite{tran2010structure}, which essentially describe which property is in between a certain subject and a certain object.
They can also be used to identify the RDF types of the subject and object, leading to the possibility of constructing a tuple that specifies which RDF type is connected to another type via a specific property.
The tool for inspecting and exploring datasets Loupe,\footnote{\url{http://loupe.linkeddata.es/loupe/}, last access 12/12/15} makes use of these triple patterns to explore the triples in a dataset.
Such result can also be achieved using a SPARQL query that retrieves the RDF types of a subject and an object as well as the connecting property between the subject and the object.
However, both of these approaches contain solely one RDF type for the subject resource, one RDF type for the object resource, and one property connecting the resources.
SLPs on the contrary may include more vocabulary terms to specify an RDF type of a resource or a property specifying a connection. 
It is a more condensed form of representation of the triple patterns and makes it easier to understand the data and faster to compute vocabulary terms recommendation.
For example, the single SLP
\begin{align*}
	(&\{\code{foaf:Person}, \code{dbo:SoccerPlayer}\}, \{\code{foaf:knows}, \code{schema:colleague}\}, \\
	 &\{\code{schema:Person}, \code{dbo:Coach}\})
\end{align*}
is enough to specify that resources of RDF types \code{foaf:Person} and \code{dbo:SoccerPlayer} are connected to resources of types \code{schema:Person} and \code{dbo:Coach} via the properties \code{foaf:knows} and \code{schema:colleague}.
One would need eight triple patterns, i.e., every combination between the RDF types and the two properties, in order to specify the relationship.
With each additional vocabulary term, the number of triples patters needed to represent the relationship rises drastically, such that it makes it harder to understand the data as well as more complicated to calculate recommendations from it.

\subsection{Vocabulary Search Engines}\label{sec:vocabSearchEngines}
Services providing a search for specific vocabulary terms generally utilize a keyword-based approach.
Their input is a string describing the desired vocabulary term, e.g., a search-string ``Person'' to find vocabulary terms describing a person.
The output is a set of RDF types and/or properties that are similar to the search-string based on some string similarity measure.
Prominent and inspiring examples of such search engines are Swoogle~\cite{DBLP:conf/cikm/DingFJPCPRDS04}, vocab.cc~\cite{vocab-cc}, Watson~\cite{d2007watson}, Falcon's concept and object Search~\cite{chengFalcons,chengObject}, and LOV~\cite{LOV}. 
Falcons contains RDF types and properties from over $4,000$ ontologies, Swoogle even from over $10,000$ ontologies, whereas LOV comprises over 500 established manually curated vocabularies, and vocab.cc provides lists of the top $100$ RDF types and properties in the Billion Triple Challenge 2011 data set.
Each service provides also various meta-information on the vocabulary terms and their vocabularies, such as the term's number of usages on the LOD cloud.
LOV also offers an API\footnote{\url{http://lov.okfn.org/dataset/lov/api}, last access 09/03/2015} that enables retrieving vocabulary terms by providing a query (e.g. ``Person'') and various other parameters, such as a \emph{type} (e.g. ``class'') or even a \emph{tag} specifying a category for a term (e.g.``people'').
The results are ordered based on a sophisticated ranking method adapting term frequency inverse document frequency (tf-idf) to the graph-structure of vocabularies~\cite{vandenbusschealinked}.
Falcons concept search recommends further ontologies once the user selects an RDF type or a property from the result list, which can be investigated for further vocabulary terms.
Falcons object search as well as Watson let the user search for specific entities, such as ``Barack Obama'', and retrieve resources from datasets on the LOD cloud that have properties containing the search string. 
This way, they are able to suggest RDF types and properties for the retrieved resources.

Alani et al.~\cite{conf/kcap/AlaniNSSM07} propose another approach for searching ontologies from different domains. 
When searching for ontologies of a particular domain, a collection of terms that represent the given domain is retrieved and used to expand the user query.
This is especially helpful when starting to choose vocabulary terms for reuse from scratch.

As additional information on the vocabulary terms, most services exploit schema-information encoded in the vocabularies, such as sub-class, sub-property or other relations between vocabulary terms.
For example, LOV offers a SPARQL\footnote{\url{http://lov.okfn.org/dataset/lov/sparql}, last access 09/03/2015} endpoint for Linked Data practitioners and applications.
Using the endpoint, one can search for RDF types or properties that are \code{equivalent} or a \code{sub-class-of} another vocabulary term, or one can search for properties that have a specific RDF type as \code{rdfs:domain} and/or \code{rdfs:range}.
Listing~\ref{lst:lovSparql} illustrates an example query to retrieve RDF types from all vocabularies stored in LOV that are equivalent to \code{foaf:Person}.
\begin{lstlisting}[float=t, breaklines=true,language=ttl,basicstyle=\scriptsize\ttfamily, xleftmargin=.15in, label=lst:lovSparql, 
caption=\footnotesize{\textbf{SPARQL query in LOV.} Querying for RDF types (\texttt{?t}) from all vocabularies/graphs in LOV (\texttt{?src}) that are equivalent to the RDF type \codeSmall{foaf:Person}. This enables to exploit structural information encoded in the RDF vocabularies}]
PREFIX owl: <http://www.w3.org/2002/07/owl#>
PREFIX foaf: <http://xmlns.com/foaf/0.1/>

SELECT DISTINCT ?t {
  GRAPH ?src{
    ?t owl:equivalentClass foaf:Person.
}} ORDER BY ?t
\end{lstlisting}
SPARQL queries for selecting RDF types that are a \code{rdfs:subClassOf} another RDF type, or properties having a specific domain and range can be designed analogously.
The results of such queries however depend on whether the connections are defined in the vocabularies stored, i.e., T-Box specification of vocabularies.
Equivalent vocabulary terms cannot be retrieved, if vocabulary terms are not connected via links such as \code{owl:equivalentClass}, \code{rdfs:subClassOf}, \code{owl:equivalentProperty}, or others.

However, such additional information on vocabulary terms do not depict how other datasets on the LOD cloud describe their data.
TermPicker does not rely on T-Box information and the modeled connections between vocabularies, nor does it suggest vocabulary terms for \emph{specific} resources.
It rather uses the datasets on the LOD cloud to calculate schema-level patterns representing which vocabulary terms are used to describe all resources and their connections. 
This way, TermPicker's recommendations are based on the encoded A-Box specification of datasets published on the LOD cloud.
On contrary to Falcons' object search and Watson, TermPicker's input is a set of vocabulary terms, i.e., a query-SLP.
Using this input, it generates a list of other vocabulary terms that are used in conjunction with the sets contained in the query-SLP, and not a list of vocabulary terms used to describe one specific resource.
Based on this, TermPicker is able to retrieve further RDF types to describe resources of a given RDF type.
For example, for resources of type \code{foaf:Person} it can suggest the RDF types \code{schema:Person}, \code{swrc:Person}, \code{dbp:FootballPlayer}, if other LOD data providers have used these types in combination with \code{foaf:Person} to describe their data.
Summarizing, with TermPicker, connections between vocabulary terms do not have to be made explicit, but can be induced directly from datasets on the LOD cloud.

\subsection{Vocabulary Recommender Systems}\label{sec:vocRecSystems}
Existing services that recommend RDF types and properties are generally based on syntactic and semantic similarity measures as well as on algorithms that provide a statement on the popularity of a recommended term.
One prominent example is the collaborative system CORE for ontology engineering (short for: Collaborative Ontology Reuse and Evaluation)~\cite{oro28592}.
A set of initial keywords defines CORE's input.
Starting from this, CORE determines a ranked list of domain-specific ontologies considerable for reuse.
The approach uses WordNet\footnote{\url{http://wordnet.princeton.edu/}, last access 09/05/15} to expand the initial set of terms, and performs a search for each of the defined keywords on an index of ontologies.
Besides syntactic and semantic similarity measures, CORE uses manual user evaluations of suggested ontologies to raise the recommendation quality. 
A similar system was developed by Romero et al.~\cite{conf/kes/RomeroVMPP10}.
However, it measures the popularity of an ontology by the number of appearances in Wikipedia or bookmarks on Del.icio.us\footnote{\url{http://delicious.com/}, access 7/15/2014}.
The previously mentioned service Watson~\cite{d2007watson} also provides a plug-in for the NeOn ontology engineering toolkit\footnote{\url{http://www.neon-project.org/}, last access 09/05/15} that supports the engineer in reusing vocabularies.
It uses semantic information from a number of ontologies and other semantic documents published on the Web to recommend appropriate vocabulary terms.

Again, the input for these recommendation services is a single string or a set of strings specifying a vocabulary term or a domain of interest.
Whereas these services provide recommendations based on string analyzes, they do not exploit any structural information on how vocabulary terms are connected to each other.
In contrast, Falcons' Ontology Search~\cite{cheng2011empirical} provides the engineer with such information.
Compared to traditional ontology matching approach, which align ontologies based on \emph{similarity}, the authors of Falcons' Ontology Search use different kinds of relatedness, in order to identify which vocabulary terms might express similar semantics.
However, it is mainly designed to establish a general relatedness between vocabularies specifying that different vocabularies contain terms that describe similar data.
Thus, it does not investigate how data providers on the LOD cloud use vocabulary terms to describe their data and individual relations as it is done by TermPicker.

There are various tools and services that transform data from various formats, such as CSV data, into RDF, but only a few ones provide support for reusing vocabularies by integrating a vocabulary recommendation service. 
The ``data2Ontology'' module of the Datalift platform~\cite{EURECOM3707} provides suggestions to match data entities to a vocabulary term based linguistic proximity between the data entity and the vocabulary term and the quality of the vocabulary using criteria from LOV.
The data integration tool Karma~\cite{springerlink:10.1007/978-3-642-30284-8_32} contains two different types of recommending vocabulary terms.
One approach suggests so-called \emph{semantic types} for a column containing data, such as the first name of a person~\cite{krishnamurthy2015}. 
The approach analyzes the content of the column using NLP techniques and recommends an RDF type in conjunction with a datatype property containing the literal value of a column's cell. 
The other recommendation approach is based on what the user has previously modeled~\cite{taheriyan16:jws}.
For example, if she has already modeled data entities and relationship about museum items, and the next data collection contains data on other museum items, the system is likely to recognize this and recommends the vocabulary terms that were used to model the previous data collection.
However, these tools either use string similarity, analyze the modeled data entities themselves, or rely on previously modeled data by the user, and do not consider what other data providers on the LOD cloud have used to model their data.
This way, their input for providing recommendations is very different and cannot be directly compared to TermPicker and its approach.

\subsection{Ontology Matching and Alignment}\label{sec:OntoMatch}
As schema-level patterns can be used to describe Linked Open Data on schema-level, and given that TermPicker compares the query-SLP with other SLPs calculated from existing datasets on the LOD cloud, one might consider TermPicker's approach being related to ontology matching~\cite{euzenat2007ontology}.
However, typical ontology matching techniques try to find correspondences between semantically related vocabulary terms of two or more different ontologies by applying \linebreak (semi-)automatic alignment algorithms.
In contrast, SLPs solely represent the connection between resources of specific RDF types via a set of properties.
The comparison of two SLPs is done solely syntactically, i.e., if the two sets of RDF types and the set of properties of two SLPs contain the same vocabulary terms, these two SLPs are considered the same. 
Thus, SLPs do not find any correspondences between semantically related vocabulary terms and is therefore not some type of ontology matching technique, nor can it be directly compared to such.

\section{Conclusion}\label{conclusion}
This paper presented TermPicker: a novel approach for recommending vocabulary terms for reuse.
The notion of schema-level patterns (SLPs), which are a major part of TermPicker, was introduced including the description how they are calculated from datasets on the LOD cloud.
It has been demonstrated how SLPs are used to define the SLP-feature and how Learning To Rank algorithms use the features to train a ranking model.
Two 10-fold leave-one-out evaluations were performed on the BTC 2014 and the \dyldo dataset, respectively, and the results illustrate that using the SLP-feature provide vocabulary term recommendation with a Mean Average Precision of about $\text{MAP} \approx 0.70$.
This improves the recommendation quality by about $35\%$ compared to the baselines of recommending vocabulary terms from popular vocabularies and recommending terms from the same vocabulary.
Furthermore, with a Mean Reciprocal Rank at the first five positions of $MRR@5 \approx 0.74$, the results indicate that the first relevant vocabulary term recommendation is within the first five results in $74\%$ of all queries.
Finally, based on the evaluation design that assesses the relevance of a recommendation candidate automatically (by extracting some terms from a query-SLP before using it as input for TermPicker), it seems that point-wise Learning To Rank (L2R) algorithms provide better results than pair-wise or list-wise L2R algorithms.

As future work, we intend to compare the data mining approach \emph{Association Rule Mining} to the utilized L2R algorithms.
To do so, we compose a user-study, in which the user gets recommendations based on Learning To Rank or on Association Rules for a specific assignment.
The user subsequently rates the perceived recommendation quality of both approaches, such that we can compare which approach provides the overall better recommendations.

To increase the recommendation quality based on Learning To Rank, it seems useful to consider the domain in which a vocabulary term is used most often.
As another feature, one could use the PageRank information of a given pay-level domain that uses a recommended vocabulary term.
This way, recommendations can be more domain specific.
Furthermore, each recommendation candidate can be enriched with meta-information, such as the appropriate \code{rdfs:domain} and \code{rdfs:range} information for properties, or \code{owl:equivalentClass} or other information for RDF types.
To this end, LOV provides an API or a SPARQL endpoint that can be used.

\paragraph{Acknowledgment}
We thank the staff of ZWB Kiel for providing feedback for this research. 
Specifically, Norbert Luttenberger, who, inspired by the tool \emph{ColumnPicker}, proposed the name \emph{TermPicker}, as it resembles a similar useability for picking vocabulary terms instead of columns.
Furthermore, Jesper Zedlitz provided valuable insights on distinguishing the proposed approach to existing tools and services, using the the specifications of the T-Box and the A-Box.

\balance
\bibliographystyle{abbrv}
\bibliography{paper}

\end{document}